\documentclass[11pt,a4paper,nofootinbib,superscriptaddress]{revtex4-1}

\pdfoutput=1
\usepackage[utf8]{inputenc}
\usepackage{float}
\usepackage{graphicx,bm,bbm}
\usepackage{slashed,verbatim}
\usepackage{amssymb,graphicx,epstopdf}
\usepackage{slashed,subfigure}
\usepackage{caption}
\usepackage{amsmath,mathtools}
\usepackage{epstopdf,dcolumn}
\usepackage{soul}

\allowdisplaybreaks
\usepackage[normalem]{ulem}
\usepackage{cancel}
\usepackage{xcolor}
 \usepackage[colorlinks=true,linktocpage=true,linkcolor=blue,citecolor=blue]{hyperref}
\usepackage{xcolor}
\allowdisplaybreaks

\def\be {\begin{equation}}
\def\ee {\end{equation}}
\def\bea {\begin{eqnarray}}
\def\eea {\end{eqnarray}}
\def\bc {\begin{center}}
\def\ec {\end{center}}
\def\nn {\nonumber}
\def\eps {\epsilon}

\def\mn {\mu\nu}
\def\({\left(}
\def\){\right)}
\def\[{\left[}
\def\]{\right]}
\def\sp {\shortparallel}
\newcommand \Tr{\operatorname{\text{Tr}}}

\def\sumintb{\sum\!\!\!\!\!\!\!\!\!\int\limits}
\def\sumintf{\sum\!\!\!\!\!\!\!\!\!\!\int\limits}

\def\slashed{\slash\!\!\!\!}
\def\slashedl{\slash\!\!\!}

\DeclareGraphicsExtensions{.jpg,.pdf,.eps}

\begin{document}
\title{Second-order quark number susceptibility of deconfined QCD matter in the presence of a magnetic field}

\affiliation{
	Theory Division, Saha Institute of Nuclear Physics, \\
	1/AF, Bidhannagar, Kolkata 700064, India}
\affiliation{School of Physical Sciences, National Institute of Science Education and Research, \\  Jatni, Khurda 752050, India}
\affiliation{
	Homi Bhabha National Institute, Anushaktinagar,\\
	Mumbai, Maharashtra 400094, India}

\author{Bithika Karmakar}
\email{bithika.karmakar@saha.ac.in}
\affiliation{
 	Theory Division, Saha Institute of Nuclear Physics, \\
 	1/AF, Bidhannagar, Kolkata 700064, India}
 	\affiliation{
 	Homi Bhabha National Institute, Anushaktinagar,\\
 	Mumbai, Maharashtra 400094, India}
\author{Najmul Haque}
\email{nhaque@niser.ac.in}
\affiliation{School of Physical Sciences, National Institute of Science Education and Research, \\  Jatni, Khurda 752050, India}
\affiliation{
 	Homi Bhabha National Institute, Anushaktinagar,\\
 	Mumbai, Maharashtra 400094, India}
\author{ Munshi G Mustafa}
\email{munshigolam.mustafa@saha.ac.in}
  \affiliation{
 	Theory Division, Saha Institute of Nuclear Physics, \\
 	1/AF, Bidhannagar, Kolkata 700064, India}
 	\affiliation{
 	Homi Bhabha National Institute, Anushaktinagar,\\
 	Mumbai, Maharashtra 400094, India}

\begin{abstract}{Considering the strong field approximation we compute the hard thermal loop pressure at finite temperature and chemical potential of hot and dense deconfined QCD matter in lowest Landau level in one-loop order. We consider the anisotropic pressure in the presence of the strong magnetic field {\it{i.e.}}, longitudinal and transverse pressure along parallel and perpendicular to the magnetic field direction. As a first effort we compute and discuss the anisotropic quark number susceptibility of deconfined QCD matter in lowest Landau level. The longitudinal quark number susceptibility is found  to increase with the temperature whereas the transverse one decreases with the temperature. We also compute the quark number susceptibility in the weak field approximation. We find that the thermomagnetic correction to the quark number susceptibility is very marginal in the weak field approximation.}

\end{abstract}

\maketitle

\section{Introduction}
Fluctuations of the conserved quantum numbers like the baryon number, electric charge and  strangeness number have been proposed as the probe of a hot and dense matter created in high energy heavy-ion collisions. However if one collects all the charged particles in heavy-ion collision then the net charge will be conserved and there will be no fluctuation. But all the particles can not be collected by any detector~\cite{Jeon:2000wg}. One should consider grand canonical ensemble for the case of real detector. An isolated system does not fluctuate because it is at a thermodynamic limit. But if we consider a small portion of a system which is small enough to consider the rest of the system as a bath and is large enough to ignore the quantum fluctuations then one can calculate the fluctuation of conserved quantities like baryon number using grand canonical ensemble~\cite{Asakawa:2000wh}. These fluctuations can be measured experimentally~\cite{Jeon:2000wg,Asakawa:2000wh,Koch:2001zn}. Several lattice calculations are there which calculate fluctuation and correlation of the conserved quantities~\cite{Bellwied:2015lba,Borsanyi:2013hza,Borsanyi:2012rr,Ding:2015fca,Gavai:2005yk}. The fluctuation of the conserved quantum numbers can be used to determine the degrees of freedom of the system~\cite{Asakawa:2000wh}. Second- and fourth- order quark number susceptibilities in thermal medium have been calculated using hard-thermal-loop (HTL) approximation~\cite{Haque:2014rua,Haque:2013sja,Haque:2013qta,Chakraborty:2003uw,Chakraborty:2001kx,Blaizot:2001vr} and perturbative quantum chromodynamics (pQCD)~\cite{Vuorinen:2002ue,Toimela:1984xy,Vuorinen:2003fs}. In Ref.~\cite{Haque:2018eph} the second-order quark number susceptibility (QNS), considering the finite strange-quark mass, was calculated .

On the other hand, recent findings show that the magnetic field of the order of $10^{18}$ Gauss can be created at the center of the fireball by the charged spectator particles in noncentral heavy-ion collisions~\cite{Skokov:2009qp,Kharzeev:2007jp}. The time-dependent magnetic field is created in a direction perpendicular to the reaction plane~\cite{Shovkovy:2012zn,DElia:2012ems,Fukushima:2012vr,Mueller:2014tea,Miransky:2015ava} and its strength depends on the impact parameter. The strength of the magnetic field decreases after few fm$/c$ of the collision~\cite{Skokov:2009qp}. Several activities are under way to study the properties of strongly interacting matter in the presence of a magnetic field. The effects like magnetic catalysis~\cite{Shovkovy:2012zn,Gusynin:1994xp,Gusynin:1995gt}, inverse magnetic catalysis \cite{Bali:2011qj,Ayala:2015bgv,Ayala:2014iba,Ayala:2014gwa,Bruckmann:2013oba,Farias:2016gmy,Farias:2014eca,Ferreira:2014kpa,Mueller:2015fka}, chiral magnetic effect \cite{Fukushima:2008xe,Kharzeev:2013ffa} and meson masses~\cite{Bali:2017ian,Ayala:2018zat,Das:2019ehv,Fayazbakhsh:2012vr,Coppola:2018vkw,Avancini:2015ady,GomezDumm:2017jij,Zhang:2016qrl,Avancini:2018svs,Avancini:2016fgq,Liu:2018zag,Liu:2014uwa,Liu:2016vuw,Chernodub:2011mc} in the presence of a magnetic field in noncentral heavy-ion collision have been reported. Furthermore, various thermodynamic quantities~\cite{Karmakar:2019tdp,Bandyopadhyay:2017cle}, transport coefficients~\cite{Kurian:2018dbn,Kurian:2017yxj}, dilepton production rate~\cite{Das:2019nzv,Bandyopadhyay:2016fyd,Bandyopadhyay:2017raf,Chyi:1999fc,Tuchin:2013bda,Ghosh:2018xhh}, photon production rate~\cite{Bzdak:2012fr,Basar:2012bp}, and damping of photons~\cite{Ghosh:2019kmf} in a magnetized QCD matter have been obtained.

Here for simplicity, we consider the strong ($gT<T<\sqrt{|q_fB|}$) and weak ($\sqrt{|q_fB|}<m_{th}\sim gT<T $ ) magnetic field strength with two different scale hierarchies. As a first effort in this article we, using the one-loop HTL pressure of quarks and gluons at finite quark chemical potential in the presence of a magnetic field, calculate the second-order QNS of deconfined QCD matter in these two scale hierarchies.
 
The paper is organized as follows: In Sec.~\ref{setup} we present the setup to calculate second-order QNS. In Sec.~\ref{quark_f}, the one-loop HTL free energy of quark in the presence of a strong magnetic field at finite temperature and chemical potential is calculated. The gauge boson free-energy in presence of a strong magnetic field is obtained in Sec.~\ref{gauge_boson}. In Sec.~\ref{pressure}, we discuss the anisotropic pressure and second-order QNS of the QCD matter in a strong field approximation. Considering the one-loop HTL pressure for the quark-gluon plasma in the weak-field approximation~\cite{Bandyopadhyay:2017cle}, we also calculate and discuss the second-order QNS in the presence of weak magnetic field in Sec.~\ref{wfa}. We conclude in Sec.~\ref{conclusion}. 
\section{Setup}
\label{setup}
Here we consider the deconfined QCD matter as a grand canonical ensemble. 

The free energy of the system is given by
\bea
\mathcal F(T,V,\mu)&=&U-TS-\mu N
\eea
where $\mu$ is the quark chemical potential and $U$, $N,$ and $S$ are the total energy, net quark number, and entropy of the system, respectively. Hence, the free energy density or the thermodynamic potential of the system can be written as
\bea
 F&=&\mathcal F/V=u-Ts-\mu n
\eea
where $u$, $n,$ and $s$ are the total energy density, net quark number density, and entropy density of the system, respectively. The pressure of the system is given as 
\bea
P=-F.
\eea
However, we consider the system to be anisotropic in the presence of a strong magnetic field and the free energy of the system is defined in Sec.~\ref{sfa}.

Now, the second-order QNS is defined as
 \bea
 \chi=-\frac{\partial^2 F}{\partial \mu^2}\bigg \vert_{\mu=0}=\frac{\partial^2 P}{\partial \mu^2}\bigg \vert_{\mu=0}=\frac{\partial n}{\partial \mu}\bigg \vert_{\mu=0}\label{chi_def},
 \eea
 which is the measure of the variance or the fluctuation of the net quark number. One can find out the covariance of two conserved quantities when the quark flavors have different chemical potential. Alternatively, one can work with other bases according to the system e.g., the net baryon number $\mathcal B$, net charge $\mathcal Q$ and strangeness number $\mathcal S$ or $\mathcal B$, $\mathcal Q$ and third-component of the isospin $\mathcal I_3$. In our case we take the  strangeness and charge chemical potential to be zero. Moreover, we consider same chemical potential for all flavors which results the vanishing off-diagonal quark number susceptibilities.  Thus the net second-order baryon number susceptibility is related to the second-order QNS  as $\chi_B=\frac{1}{3}\chi$.
 
The strength of the magnetic field produced in a noncentral heavy-ion collision can be up to $(10-20)m_\pi^2$ at the time of the collision~\cite{Bzdak:2011yy}. However, it decreases very fast being inversely proportional to the square of time~\cite{Bzdak:2012fr,McLerran:2013hla}. But if one considers finite electric conductivity of the medium, then the magnetic field strength will not die out very fast~\cite{Tuchin:2013bda,Tuchin:2012mf,Tuchin:2013ie,Tuchin:2013apa, Steinert:2013fza,Hattori:2016cnt,Ghosh:2019ubc,Fukushima:2017lvb}. We consider two different cases with strong and weak magnetic field in this article.
\section{Strong magnetic field}
\label{sfa}
In this section we consider the strong field scale hierarchy $gT < T < \sqrt{eB}$. In the presence of the magnetic field, the energy of charged fermion becomes $E_n=\sqrt{k_3^2+m_f^2+2n q_fB}$ where $k_3$ is the momentum of a fermion along the magnetic field direction, $m_f$ is the mass of the fermion and the Landau level, $n$, can vary from 0 to $\infty$. The transverse momentum of the fermion becomes quantized. It can be shown that at very high value of the magnetic field, the contribution from all the Landau levels, except the lowest Landau level, can be ignored~\cite{Bandyopadhyay:2016fyd}. Consequently, the dynamics becomes $(1+1)$ dimensional when one considers only the lowest Landau level (LLL). The general structures of the quark and gluon self-energy in the presence of the magnetic field have been formulated in Ref.~\cite{Karmakar:2019tdp} at finite temperature but for zero quark chemical potential. Here we extend it for the case of nonzero quark chemical potential. In the presence of the strong magnetic field, the general structure of quark self-energy can be written from Ref~\cite{Karmakar:2019tdp} as
\begin{center}
\begin{figure}[tbh!]
 \begin{center}
 \includegraphics[scale=0.6]{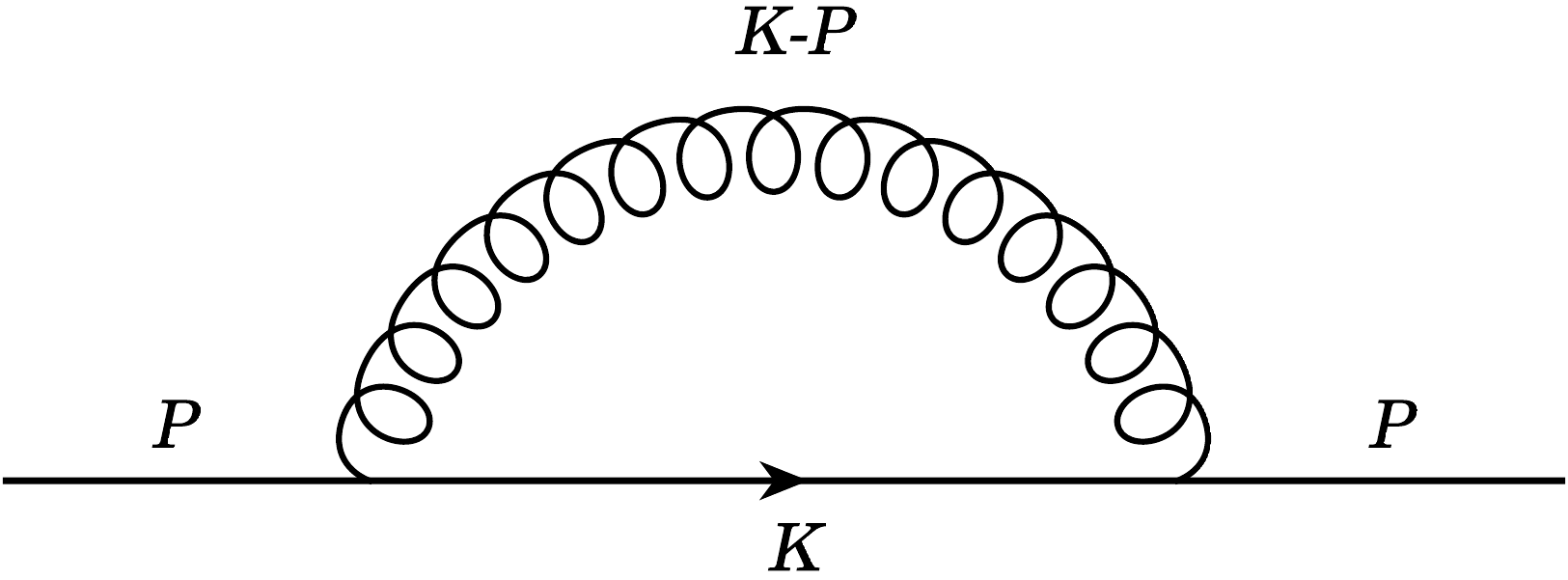} 
 \caption{Quark self-energy diagram}
  \label{quark_se}
 \end{center}
\end{figure}
\end{center} 
\bea
\Sigma(p_0,p_3)&=& a \slashedl u + b \slashedl n + c\gamma_5 \slashedl u +d\gamma_5 \slashedl n ,
\eea
where the rest frame of heat bath velocity is $u_\mu=(1,0,0,0)$ and the direction of magnetic field is $n_\mu=(0,0,0,1)$. Now, the various form factors can be obtained as
\bea
a&=&\frac{1}{4}\Tr[\Sigma \slashedl u], \label{a_def}\\
b &=& -\frac{1}{4}\Tr[\Sigma \slashedl n] , \label{b_def}\\
c&= &\frac{1}{4}\Tr[ \gamma_5 \Sigma\slashedl u], \label{c_def}\\
d&=&-\frac{1}{4}\Tr[ \gamma_5 \Sigma\slashedl n].\label{d_def}
\eea 
Considering Fig.~\ref{quark_se} the above form factors are calculated up to $\mathcal O[(\mu/T)^4]$ in Appendix~\ref{quark_ff} as
\bea
a=-d
&=&c_1 \Bigg[ \frac{p_0}{p_0^2-p_3^2}c_2 +\frac{(p_0^2+p_3^2)}{2(p_0^2-p_3^2)^2}c_3\Bigg]\label{ff_a},\\
b=-c
&=&-c_1\Bigg[ \frac{p_3}{p_0^2-p_3^2} c_2 +\frac{p_0p_3}{(p_0^2-p_3^2)^2}c_3\Bigg]\label{ff_b},
\eea
where $c_1,c_2,$ and $c_3$ are defined in Eqs.~\eqref{c1c2c3}.

\subsection{One-loop quark free energy in the presence of a strongly magnetized medium}
\label{quark_f}
In this section we calculate the quark free energy within the HTL approximation using the form factors of quark self-energy in~\eqref{ff_a} and~\eqref{ff_b}. The quark free energy can be written as
\bea
F_q=- d_F \sumintf_{\{p_0\}}~\frac{d^3p}{(2\pi)^3}\ln{\left(\det[ S^{-1}_{\text{eff}}(p_0, p_3)]\right)} ,
\label{fe}
\eea
where $d_F=N_c N_f$. Here we use the sum-integral as
\bea
\sumintf_{\{p_0\}} \equiv T\!\!\!\!\!\!\sum_{p_0=(2n+1)\pi i T +\mu} \int \frac{d^3p}{(2\pi)^3}.
\eea

The inverse of the effective fermion propagator can be written as
\bea
 S^{-1}_{\text{eff}}=\slashed{P_{\sp}}+\Sigma &=& (p_0+a)\slashedl{u}+(b-p_3)\slashedl{n}+c \gamma_5 \slashedl{u}+d \gamma_5 \slashedl{n}\nn\\
&=&(p_0+a)\gamma^0+(b-p_3)\gamma^3+c \gamma_5 \gamma^0+d \gamma_5 \gamma^3.
\eea
Now we evaluate the determinant as
\bea
\det[ S^{-1}_{\text{eff}}]&=&\bigg((b+c-p_3)^2-(a+d+p_0)^2\bigg)\bigg((-b+c+p_3)^2-(a-d+p_0)^2\bigg)\nn\\
&=&(p_0^2-p_3^2)\bigg((p_0+2a)^2-(p_3-2b)^2\bigg)\nn\\
&=& P_{\sp}^2\(P_\shortparallel^2+4 a p_0+4 b p_3+4 a^2- 4b^2\)\nn\\
&=&  P_{\sp}^4 \left(1+\frac{4a^2-4b^2+4ap_0+4bp_3}{P_{\sp}^2}\right),
\eea
where we have used $d=-a$ and $c=-b$. 

So Eq.~\eqref{fe} becomes
\bea
F_q&=&- d_F \sumintf_{\{p_0\}}~\frac{d^3p}{(2\pi)^3} \ln\bigg[P_{\sp}^4\(1+\frac{4a^2-4b^2+4ap_0+4bp_3}{P_{\sp}^2}\)\bigg]\nn\\
&=& -2 d_F  \sumintf_{\{p_0\}}~\frac{d^3p}{(2\pi)^3}\ln{(-P_{\sp}^2)}- d_F \sumintf_{\{p_0\}}~\frac{d^3p}{(2\pi)^3}
\ln\bigg[1+\frac{4a^2-4b^2+4ap_0+4bp_3}{P_{\sp}^2}\bigg]\nn\\
&=& F^{\text{ideal}}_q+ F'_q,
\eea
where the free energy of free quarks in the presence of a magnetic field~\cite{Strickland:2012vu} reads as
\bea
F^{\text{ideal}}_q&=& -2 d_F  \sumintf_{\{p_0\}}~\frac{d^3p}{(2\pi)^3}\ln{(-P_{\sp}^2)}
= -2 d_F \sum_f \frac{q_f B}{(2 \pi)^2} \sumintf_{\{p_0\}} dp_3 \ln{(-P_{\sp}^2)}\nn\\
&=&- d_F \sum_f \frac{q_f B T^2}{6}\bigg( 1+12\hat \mu^2 \bigg),
\eea
where $\hat \mu=\mu/2\pi T$, and
\bea
F'_q &=&- d_F \sumintf_{\{p_0\}}~\frac{d^3p}{(2\pi)^3}
\ln\bigg[1+\frac{4a^2-4b^2+4ap_0+4bp_3}{P_{\sp}^2}\bigg]\nn\\ 
&=&- d_F \sumintf_{\{p_0\}}~\frac{d^3p}{(2\pi)^3}\bigg[\frac{4\(a p_0 +b p_3\)}{P_{\sp}^2}+
\frac{4\(a^2 P_{\sp}^2-b^2P_{\sp}^2-2a^2 p_0^2-2b^2 p_3^2-4ab p_0p_3\)}{P_{\sp}^4}\nn\\
&&\hspace{4cm}+ \mathcal{O}(g^6)\bigg]\label{F_2_exp},
\eea
where we have kept terms up to $\mathcal{O}(g^4)$ to obtain the analytic expression of free energy. The expansion made above is valid for $g^2 (q_fB/T^2) < 1$, which can be
realized as  $(q_fB)/T^2 \gtrsim 1$ and $g \ll 1$.

As the fermions are considered to be in LLL in the strong field approximation, Eq.~(\ref{F_2_exp}) becomes
\bea
F'_q&=& - d_F \sum_f\frac{q_f B}{(2\pi)^2}\sumintf_{\{p_0\}}~dp_3~\bigg[\frac{4\(a p_0 +b p_3\)}{P_{\sp}^2}+
\frac{4\(a^2 P_{\sp}^2-b^2P_{\sp}^2-2a^2 p_0^2-2b^2 p_3^2-4ab p_0p_3\)}{P_{\sp}^4}\nn\\
&&\hspace{4.5cm}+ \mathcal{O}(g^6)\bigg].
\label{Fq'_ini}
\eea
The sum-integrals are calculated in Appendix~\ref{quark_free_energy} and the expression for the quark free energy up to $\mathcal{O}(g^4)$ is obtained by adding individual contributions as 
\bea
F_q&=&F_q^{\text{ideal}}+F_q'=- d_F \sum_f \frac{q_f B T^2}{6}\Big(1+\frac{ 3\mu^2}{4\pi^2T^2} \Big)\\
&+&4 d_F \sum_f \frac{g^2 C_F (q_f B)^2}{(2\pi)^4}\(\frac{\Lambda}{4\pi T}\)^{2\eps}\Bigg[\frac{1}{\eps}\bigg(-\frac{1}{2}\ln 2+\frac{7\mu^2 \zeta(3)}{16\pi^2 T^2}-\frac{31\mu^4 \zeta(5)}{64\pi^4T^4} \bigg)-\frac{3\gamma_E \ln 2}{2}\nn\\
&+&\ln 2 \ln \pi-\frac{1}{2}\ln 2 \ln 16\pi +\frac{g^2 C_F (q_f B)}{4 \pi^2}\frac{63\ln2^2 \zeta(3)}{72\pi^2T^2}-\frac{g^2 C_F (q_f B)}{4 \pi^2}\frac{217(q_fB)^2\zeta(5)}{36864\pi^4T^6}\nn\\
&\times&\Big(\gamma_E+2\ln2-12\ln G \Big)^2+\frac{\mu^2}{1152\pi^2T^4}\bigg\{ \frac{7\zeta(3) g^2 C_F (q_f B)}{4 \pi^2}\Big( 3+3\gamma_E+4\ln 2 -36\ln G\Big)^2\nn\\
&+&504T^2\zeta(3)\Big(3\gamma_E+8\ln 2-\ln \pi \Big)-\frac{36 \ln 2}{\pi^2}\frac{g^2 C_F (q_f B)}{4 \pi^2}\Big( 49\zeta(3)^2+186\ln 2 \zeta(5)\Big)\bigg\}\nn\\
&-&\frac{7 (q_fB)^2 \mu^2}{737280\pi^4T^8}\frac{g^2 C_F (q_f B)}{4 \pi^2}\bigg\{-31\zeta(5)\Big( -15+15\gamma_E+16\ln 2\Big)\Big( \gamma_E+2\ln 2 -12\ln G\Big)\nn\\
&-&\frac{48825}{\pi^4}\zeta(3)^2\zeta(5)-\frac{9525\zeta(7)}{\pi^2}\Big(\gamma_E+2\ln 2-12\ln G \Big)^2+55800\zeta(5)\zeta'(-3)\nn\\
&\times&\Big(\gamma_E+2\ln2-12\ln G \Big) \bigg\}+\frac{\mu^4}{69120\pi^6 T^6}\bigg\{-1080\pi^2T^2\bigg(98\zeta(3)^2+31\zeta(5)\Big( 3\gamma_E+8\ln 2 -\ln \pi\Big) \bigg)\nn\\
&-&\frac{g^2 C_F (q_f B)}{4 \pi^2}\bigg(14\pi^4 \zeta(3)\Big( 15\gamma_E+16\ln 2\Big)\Big( 3+3\gamma_E+4\ln 2-36\ln G\Big)-46305\zeta(3)^3\nn\\
&+&2790\pi^2 \zeta(5)\Big( 3+3\gamma_E-4\ln 2-36\ln G\Big)^2-820260 \ln 2 \zeta(3) \zeta(5)-1028700\ln 2^2 \zeta(7)\nn\\
&-&25200\pi^4 \zeta(3) \zeta'(-3)\Big(3+3\gamma_E+4\ln 2-36\ln G \Big) \bigg) \bigg\}-\frac{ \mu^4 (q_fB)^2}{5308416\pi^4 T^{10}}\frac{g^2 C_F (q_f B)}{4 \pi^2}\nn\\
&\times&\bigg\{ \frac{10897740}{\pi^6}\zeta(3)\zeta(5)^2+\frac{37804725}{\pi^6}\zeta(3)^2 \zeta(7)+\frac{2253510\zeta(9)}{\pi^4}\Big( \gamma_E+2\ln 2-12\ln G\Big)^2\nn\\
&+&\frac{24003}{\pi^2}\zeta(7)\Big( \gamma_E+2\ln 2-12\ln G\Big)\Big(-15+15\gamma_E+16\ln 2-1800\zeta'(-3) \Big)\nn\\
&+&\frac{31\zeta(5)}{100}\bigg(14175-40950\gamma_E+26775\gamma_E^2+68240\gamma_E \ln 2+41728\ln 2^2+151200\ln G\nn\\
&-&151200\gamma_E \ln G-240\ln 2 \Big( 231+640\ln G\Big)+3175200\zeta'(-5)\Big( \gamma_E+2\ln 2-12\ln G\Big)\nn\\
&-&226800\zeta'(-3)\Big( -15+15\gamma_E+16\ln 2\Big)+204120000\zeta'(-3)^2\bigg)\bigg\}\Bigg].
\eea

The divergences are regulated by adding suitable counterterms as
\bea
F_{ct}&=&-4 d_F \sum_f \frac{g^2 C_F (q_f B)^2}{(4 \pi^2)^2}\Bigg[  -\frac{1}{2\eps}\bigg(\ln 2 -\frac{7}{8\pi^2}\frac{\mu^2}{T^2}  \zeta(3)+\frac{31 }{24}\frac{\mu^4}{T^4} \frac{3}{4\pi^4}\zeta(5)\bigg)\Bigg].
\eea

The 
 renormalized quark free-energy is given as
\bea
F_q^r
&=&- d_F \sum_f\frac{q_f B T^2}{6}\Big(1+12\hat \mu^2 \Big)+4 d_F \sum_f \frac{g^2 C_F (q_f B)^2}{(2\pi)^4}\Bigg[-\ln 2 \ln \frac{\hat \Lambda}{2}-\frac{3\gamma_E \ln 2}{2}\nn\\
&+&\ln 2 \ln \pi-\frac{1}{2}\ln 2 \ln 16\pi +\frac{g^2 C_F (q_f B)}{4 \pi^2}\frac{63\ln2^2 \zeta(3)}{72\pi^2T^2}-\frac{g^2 C_F (q_f B)}{4 \pi^2}\frac{217(q_fB)^2\zeta(5)}{36864\pi^4T^6}\nn\\
&\times&\Big(\gamma_E+2\ln2-12\ln G \Big)^2+\frac{7\hat \mu^2}{2}\zeta(3) \ln \frac{\hat \Lambda}{2}+\frac{\hat \mu^2}{288T^2}\bigg\{ \frac{7\zeta(3) g^2 C_F (q_f B)}{4 \pi^2}\nn\\
&\times&\Big( 3+3\gamma_E+4\ln 2 -36\ln G\Big)^2+504T^2\zeta(3)\Big(3\gamma_E+8\ln 2-\ln \pi \Big)-\frac{36 \ln 2}{\pi^2}\frac{g^2 C_F (q_f B)}{4 \pi^2}\nn\\
&\times&\Big( 49\zeta(3)^2+186\ln 2 \zeta(5)\Big)\bigg\}-\frac{7 (q_fB)^2 \hat \mu^2}{184320\pi^2T^6}\frac{g^2 C_F (q_f B)}{4 \pi^2}\bigg\{-31\zeta(5)\Big( -15+15\gamma_E+16\ln 2\Big)\nn\\
&\times&\Big( \gamma_E+2\ln 2 -12\ln G\Big)-\frac{48825}{\pi^4}\zeta(3)^2\zeta(5)-\frac{9525\zeta(7)}{\pi^2}\Big(\gamma_E+2\ln 2-12\ln G \Big)^2\nn\\
&+&55800\zeta(5)\zeta'(-3)\Big(\gamma_E+2\ln2-12\ln G \Big) \bigg\}-\frac{31\hat \mu^4}{2}\zeta(5) \ln \frac{\hat \Lambda}{2}+\frac{\hat \mu^4}{4320\pi^2 T^2}\bigg\{-1080\pi^2T^2\nn\\
&\times&\bigg(98\zeta(3)^2+31\zeta(5)\Big( 3\gamma_E+8\ln 2 -\ln \pi\Big) \bigg)-\frac{g^2 C_F (q_f B)}{4 \pi^2}\bigg(14\pi^4 \zeta(3)\Big( 15\gamma_E+16\ln 2\Big)\nn\\
&\times&\Big( 3+3\gamma_E+4\ln 2-36\ln G\Big)-46305\zeta(3)^3+2790\pi^2 \zeta(5)\Big( 3+3\gamma_E-4\ln 2-36\ln G\Big)^2\nn\\
&-&820260 \ln 2 \zeta(3) \zeta(5)-1028700\ln 2^2 \zeta(7)-25200\pi^4 \zeta(3) \zeta'(-3)\Big(3+3\gamma_E+4\ln 2-36\ln G \Big) \bigg) \bigg\}\nn\\
&-&\frac{ \hat \mu^4 (q_fB)^2}{331776 T^{6}}\frac{g^2 C_F (q_f B)}{4 \pi^2}\bigg\{ \frac{10897740}{\pi^6}\zeta(3)\zeta(5)^2+\frac{37804725}{\pi^6}\zeta(3)^2 \zeta(7)+\frac{2253510\zeta(9)}{\pi^4}\nn\\
&\times&\Big( \gamma_E+2\ln 2-12\ln G\Big)^2+\frac{24003}{\pi^2}\zeta(7)\Big( \gamma_E+2\ln 2-12\ln G\Big)\Big(-15+15\gamma_E+16\ln 2\nn\\
&-&1800\zeta'(-3) \Big)+\frac{31\zeta(5)}{100}\bigg(14175-40950\gamma_E+26775\gamma_E^2+68240\gamma_E \ln 2+41728\ln 2^2\nn\\
&+&151200\ln G-151200\gamma_E \ln G-240\ln 2 \Big( 231+640\ln G\Big)+3175200\zeta'(-5)\nn\\
&\times&\Big( \gamma_E+2\ln 2-12\ln G\Big)-226800\zeta'(-3)\Big( -15+15\gamma_E+16\ln 2\Big)+204120000\zeta'(-3)^2\bigg)\bigg\}\Bigg]
\eea
where $\hat \Lambda=\Lambda/2\pi T$, $\hat \mu=\mu/2\pi T$, $G\approx1.2824$ is Glaisher's constant and $\gamma_E\approx0.5772$ is Euler-Mascheroni constant.
\subsection{Gauge boson free energy in a strongly magnetized medium}
\label{gauge_boson}
The general structure of gauge boson self-energy can be written from Ref.~\cite{Karmakar:2018aig} as
\bea
\Pi^{\mn}=\alpha B^{\mn}+ \beta R^{\mn}+\gamma Q^{\mn}+\delta N^{\mn},
\eea
where $\alpha$, $\beta$, $\gamma$ and $\delta$ are the form factors. $B^{\mn}$, $R^{\mn}$, $Q^{\mn}$ and $N^{\mn}$ are the basis tensors of gluon self-energy. The form factors are calculated in Ref.~\cite{Karmakar:2018aig} for zero quark chemical potential. Here we extend the calculation for nonzero quark chemical potential. The effect of nonzero quark chemical potential is reflected only in the Debye mass because the quark loop gets modified. But in the presence of a strong magnetic field the Debye mass does not change due to dimensional reduction. The form factors can be calculated as
\bea
\alpha &=&B^{\mn}\Pi_{\mn}=\frac{m_D^2}{\bar u^2}\left[1-\mathcal{T}_P(p_0,p)\right]-\sum_f \frac{(\delta m_{D,f}^2)_s}{\bar u^2}e^{{-p_\perp^2}/{2q_fB}}~\frac{p_3^2}{p_0^2-p_3^2}, \label{b_sf} \\
\beta&=&R^{\mn}\Pi_{\mn}=\frac{m_D^2}{2}\left[\frac{p_0^2}{p^2}-\frac{P^2}{p^2}\mathcal{T}_P(p_0,p)\right] , \label{c_sf} \\
\gamma&=&Q^{\mn}\Pi_{\mn}= \frac{m_D^2}{2}\left[\frac{p_0^2}{p^2}-\frac{P^2}{p^2}\mathcal{T}_P(p_0,p)\right]+\sum_f \frac{(\delta m_{D,f}^2)_s}{\bar u^2} e^{{-p_\perp^2}/{2q_fB}}~ \frac{p_3^2}{p_0^2-p_3^2}, \label{d_sf}\\
\delta&=&\frac{1}{2}N^{\mn}\Pi_{\mn}=\sum_f  (\delta m_{D,f}^2)_s\frac{\sqrt{\bar n^2}}{\sqrt{\bar u^2}}~ e^{{-p_\perp^2}/{2  eB}}\frac{p_0p_3}{p_0^2-p_3^2}, \label{a_sf}
\eea
where ${\bar u}^2 = - p^2/P^2$, ${\bar n}^2 = -p_{\perp}^2/p^2$ and
\bea \mathcal{T}_P(p_0,p)=\frac{p_0}{2p}\ln\frac{p_0+p}{p_0+p}.
\eea
The thermal and magnetic correction of the Debye screening mass is given as
\bea
m_D^2&=&\frac{g^2N_c T^2}{3},\\
(\delta m_{D,f}^2)_s&=&\frac{g^2|q_fB|}{2\pi T}\int\limits_{-\infty}^\infty\frac{dk_3}{4\pi}~ \bigg[n_F(k_3+\mu)\Big\{1-n_F(k_3+\mu)\Big\}+n_F(k_3-\mu)\Big\{1-n_F(k_3-\mu)\Big \}\bigg]\nn\\
&=& \frac{g^2 |q_fB|}{4\pi^2},\\
(m^s_D)^2&=&m_D^2 +\sum_f (\delta m_{D,f}^2)_s=m_D^2+(\delta m_D^2)_s.
\eea
The total  gluon free-energy expanded up to ${\mathcal O}[g^4]$ is given by
\bea
F_g 
&\approx&d_A\left[\sumintb_{P} ~\ln\left(-P^2\right) -\frac{\alpha +\beta +\gamma}{2P^2}-\frac{\alpha^2+\beta^2+\gamma^2+2\delta^2}{4P^4}\right] \label{Fsg_expan}
\eea 
where $d_A=N_c^2-1$.

The gluon free energy is calculated in details in Ref.~\cite{Karmakar:2019tdp}. Here we give the final expression. The renormalized total gluon free energy containing both hard and soft contributions is given as
 \bea
F_g^r&=& \frac{d_A}{(4\pi)^2}\Bigg[- \frac{16 \pi^4 T^4}{45}+\frac{2C_A g^2\pi^2T^4}{9}+\frac{1}{12}\(\frac{C_A g^2T^2}{3}\)^2 \left(8 - 3 \gamma_E - \pi^2 + 4\ln 2 - 3 \ln\frac{\hat \Lambda}{2}\right)\nn\\
 &+&
 \frac{2N_f\pi ^2 T^2}{9} \(\frac{g^2}{4\pi^2}\)^2  \sum_{f} q_fB\bigg( 36\ln G-4+3\ln\hat \Lambda\bigg) +\left(N_f^2+\sum_{f_1,f_2}\frac{q_{f_1}B}{q_{f_2}B}
 \right)\nn\\
 &\times& \frac{g^4T^4}{32} \left(-\frac{12 \zeta '(4)}{\pi ^4}+\frac{2}{15} \left(\ln \frac{\hat \Lambda
 }{2}+\gamma_E +\ln 4 \pi \right)-\frac{17}{75}\right) -\frac{1}{2}\(\frac{g^2}{4\pi^2}\)^2\nn\\
 &\times&	\sum_{f_1,f_2} q_{f_1}B q_{f_2}B \bigg( 4-4 (\ln 2-1) (\ln \hat \Lambda +\gamma_E )-\frac{\pi ^2}{3}+2 (\ln
   2-2) \ln 2\bigg)
 	- \frac{C_AN_f g^4T^4}{36} \nn\\
 	&\times&\bigg(3 - 24\ln G - 2 \ln\frac{\hat \Lambda}{2}\bigg)-\sum_f \frac{C_Ag^4T^2q_fB}{144\pi^2} \bigg(\pi ^2-4+12\ln \frac{\hat \Lambda }{2}-  2\ln 2\Big(6\gamma_E+4\nn\\
 	&+&3\ln2-6\ln \frac{\hat \Lambda }{2}\Big) +12 \gamma_E  \bigg)\Bigg]-\frac{d_A( m_D^s)^3T}{12\pi}\label{fg}
 	\eea
 	where $C_A=3$ is the color factor which is associated with gluon emission from a gluon.
 \subsection{QCD coupling constant}
In the strong magnetic field region we use the QCD coupling constant obtained in Ref.~\cite{Ayala:2018wux} which depends on both momentum transfer and magnetic field
 \bea
\alpha_s(\Lambda^2,|eB|)=\frac{\alpha_s(\Lambda^2)}{1+b_1\alpha_s(\Lambda^2) \ln\Big(\frac{\Lambda^2}{\Lambda^2+|eB|} \Big)} 	
 	\eea
 where the one-loop running coupling in the absence of magnetic field is given by
 \bea
 \alpha_s(\Lambda^2)=\frac{1}{b_1\ln\Big( \Lambda^2/\Lambda_{\overline{MS}^2}\Big)},\label{alpha_thermal}
 \eea
 with $b_1=\frac{11N_c-2N_f}{12\pi}$ and $\Lambda_{\overline{MS}}=176$ MeV~\cite{Beringer:1900zz} at $\alpha_s(1.5 GeV)=0.326$ for $N_f=3$. 	
 In Sec.~\ref{wfa} we use the coupling constant defined in Eq.~\eqref{alpha_thermal} which is independent of the magnetic field as the corresponding magnetic field is feeble.
\subsection{Longitudinal and transverse pressure and corresponding susceptibilities}
\label{pressure}
Free energy density of the quark-gluon plasma in the presence of a strong magnetic field is given by 
\bea
F=u-Ts-\mu n -eB\cdot M\label{F_sfa},
\eea
where $u$ is total the energy density 
and magnetization per unit volume is given by
\bea
M=-\frac{\partial F}{\partial (eB)}. \label{magnetization}
\eea
The pressure becomes anisotropic~\cite{Karmakar:2019tdp,PerezMartinez:2007kw} due to the magnetization acquired by the system in presence of strong magnetic field which results in two different pressure along parallel and perpendicular to the magnetic field direction. The longitudinal pressure is given as
\bea
P_z=-F =-(F_q^r+F_g^r).
\eea
and transverse pressure is given as
\bea
P_{\perp}=-F-eB \cdot M.
\eea

\begin{center}
\begin{figure}[tbh!]
 \begin{center}
 \includegraphics[scale=0.49]{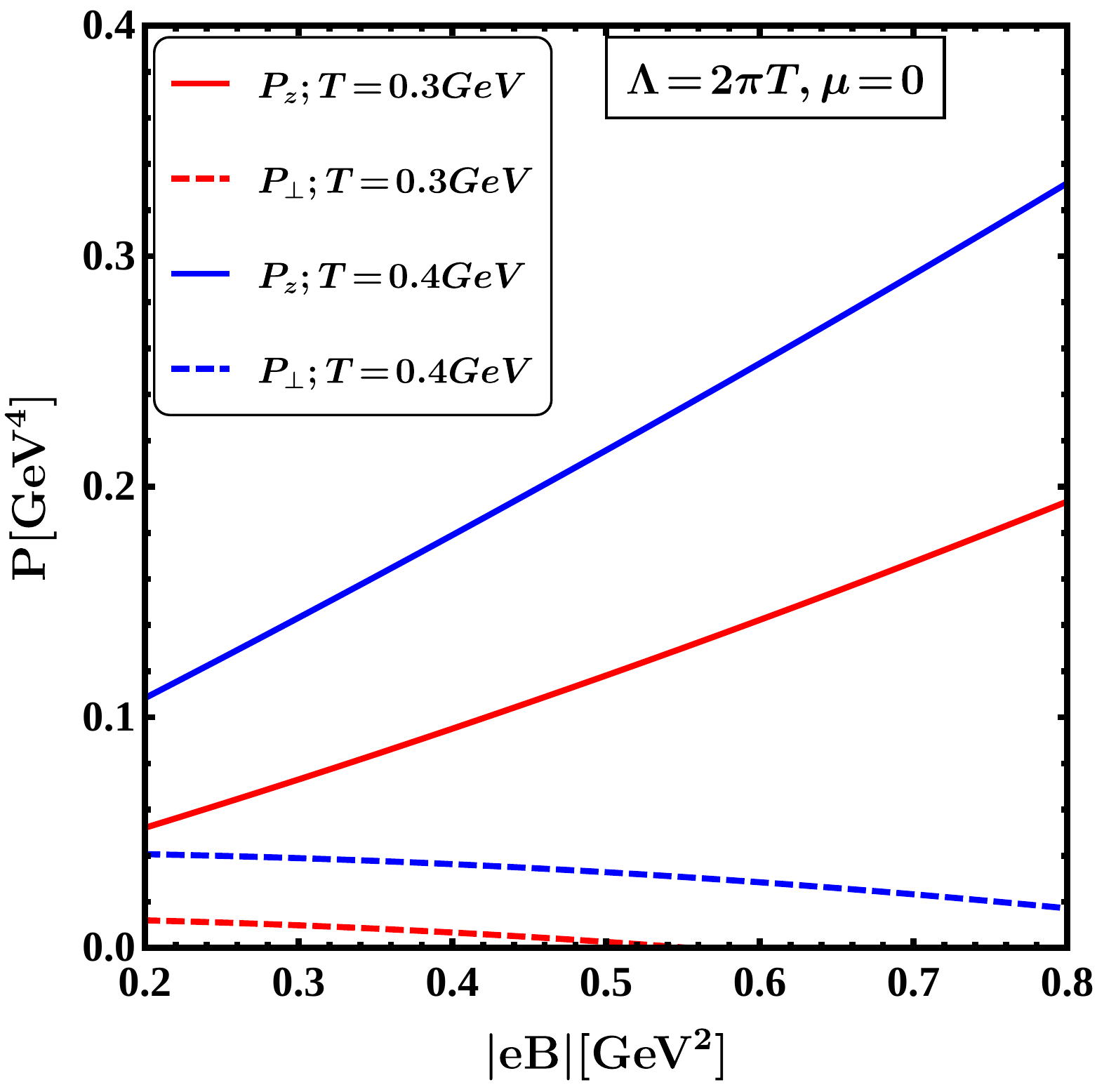} 
  \includegraphics[scale=0.49]{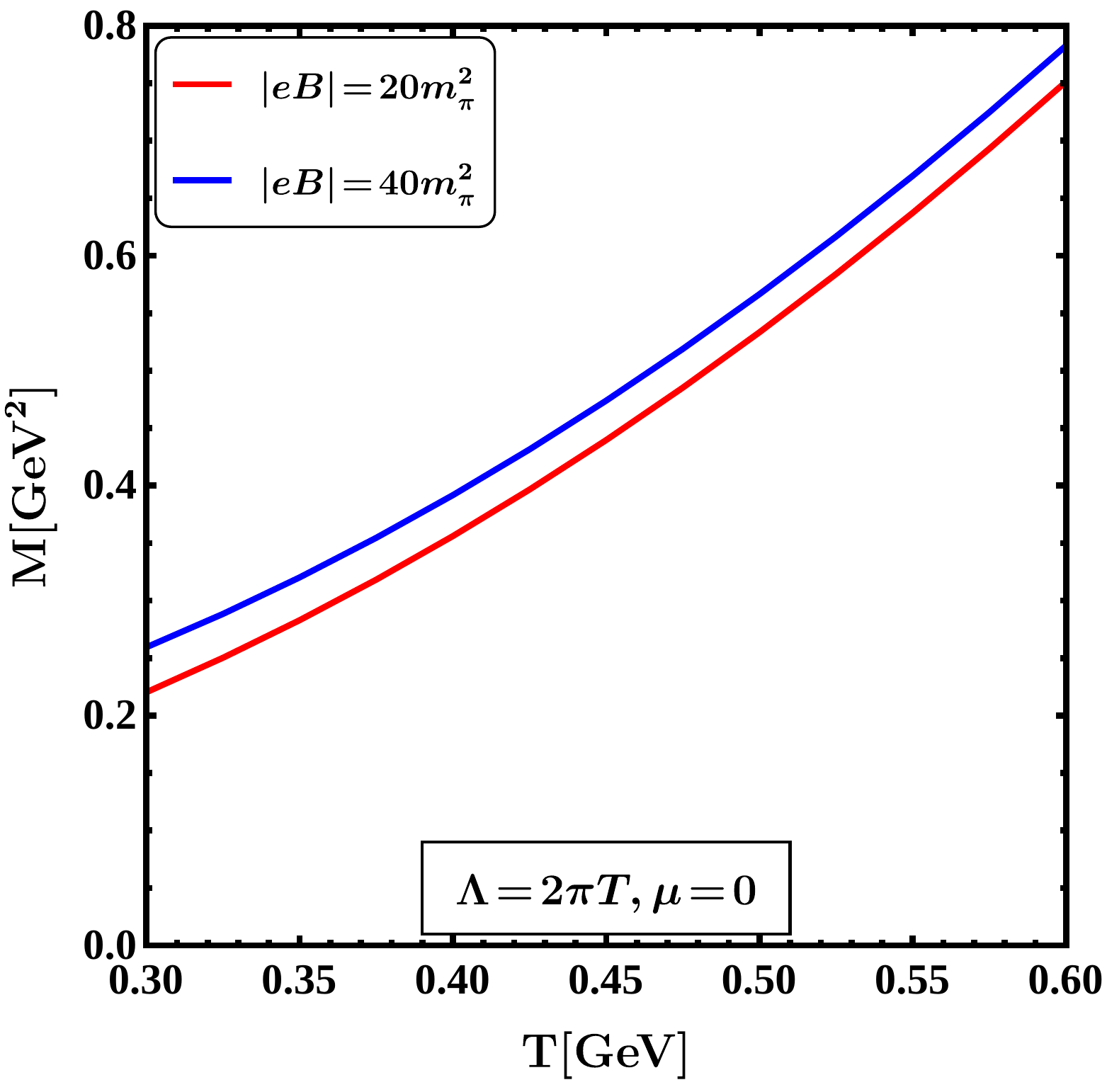} 
 \caption{Variation of the longitudinal and transverse pressure at $\mu=0$ with magnetic field is shown in left panel. Magnetization as a function of temperature at $\mu=0$ is shown in right panel for $N_f=3$.}
  \label{P_mag_sfa}
 \end{center}
\end{figure}
\end{center} 

In the left panel of Fig.~\ref{P_mag_sfa} we show the variation of longitudinal and transverse pressure with the strength of the magnetic field at $\mu=0$. It can be seen that the longitudinal pressure (pressure along the magnetic field direction) of magnetized quark-gluon plasma (QGP) increases with the magnetic field whereas the transverse pressure is opposite in nature. This indicates that the system may elongate along the longitudinal direction and compress along the transverse direction at a high magnetic field. In the right panel of Fig.~\ref{P_mag_sfa} magnetization of the system is plotted with the temperature. The positive value of the magnetization implies paramagnetism of the strongly magnetized QCD medium which is also observed in recent lattice calculation~\cite{Bali:2014kia}. It is noted that the pressure and magnetization plots of Fig.~\ref{P_mag_sfa} shows qualitative matching with the lattice result of Ref.~\cite{Bali:2014kia}. However, quantitatively the results are different due to the fact that one only gets correct perturbative coefficients of $g^0$ and $g^3$ in leading order of HTLpt. Thus one should go beyond one loop to get complete result up to $\mathcal O(g^5)$.

One gets two different second-order QNS, namely,  along the longitudinal ($\chi_z$) and transverse ($\chi_{\perp}$) direction in the presence of a strong magnetic field.
The longitudinal second-order QNS can be obtained as
\bea
\chi_z= \frac{\partial^2P_z}{\partial \mu^2}\bigg \vert_{\mu=0},
\eea
whereas the transverse  one can be obtained as
\bea
\chi_\perp= \frac{\partial^2P_\perp}{\partial \mu^2}\bigg \vert_{\mu=0}.
\eea
The longitudinal pressure of noninteracting quark-gluon gas in the presence of strong magnetic field is given as
\bea
P_{sf}= \sum_f N_c N_f \,q_fB \frac{T^2}{6}(1+12\hat \mu^2)+(N_c^2-1)\frac{\pi^2 T^4}{45}.
\eea
The second-order longitudinal QNS for the ideal quark gluon plasma is given as
\bea
\chi_{sf}=\sum_f N_c N_f \frac{q_fB}{\pi^2}.
\label{chi_sfa_ideal}
\eea

The transverse pressure of ideal quark-gluon plasma is given as
\bea
P^{\perp}_{sf}=(N_c^2-1)\frac{\pi^2 T^4}{45}.
\eea
Thus, the second-order transverse QNS of the ideal quark-gluon plasma vanishes.

\begin{center}
\begin{figure}[tbh!]
 \begin{center}
 \includegraphics[scale=0.5]{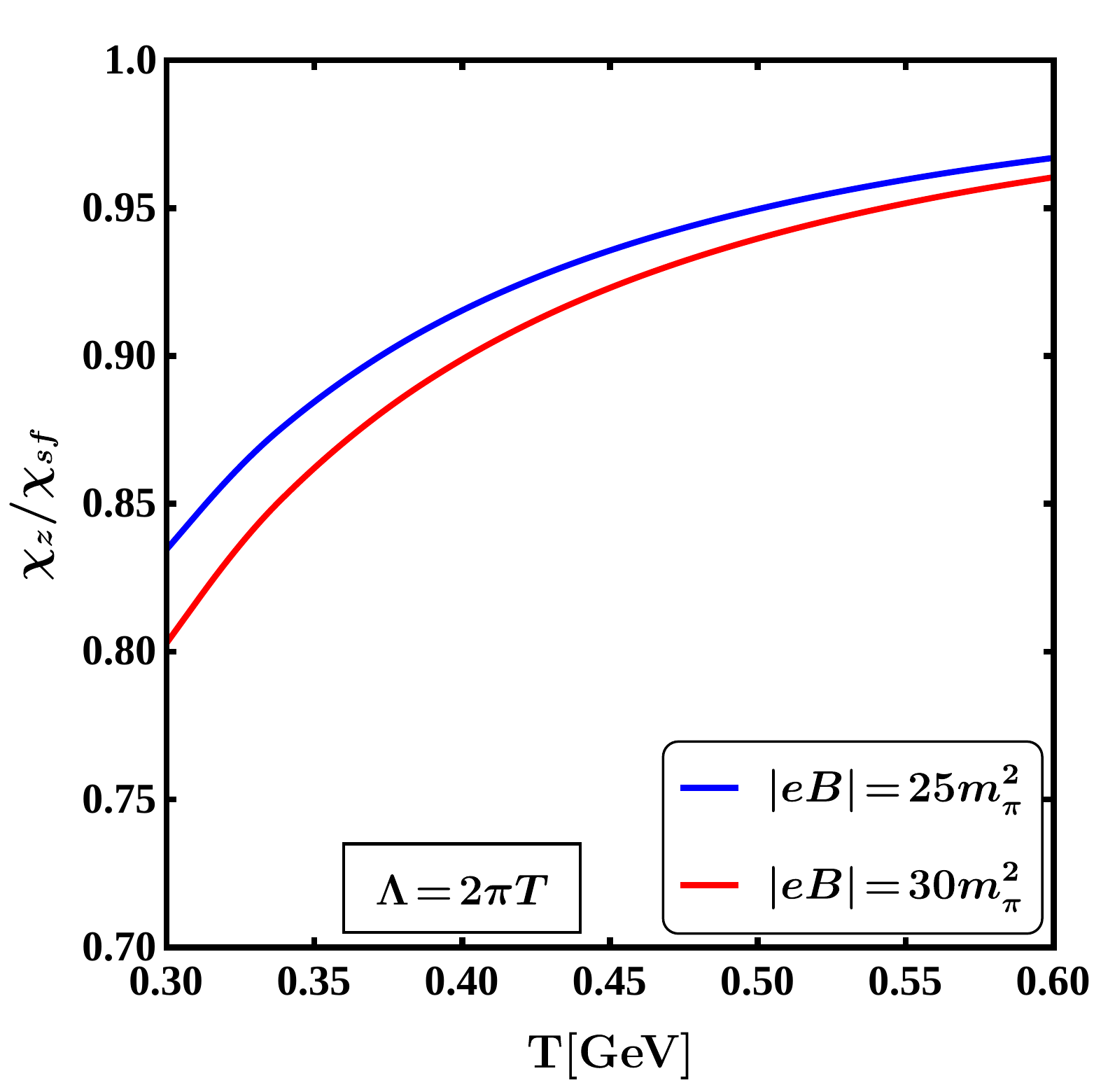} 
  \includegraphics[scale=0.49]{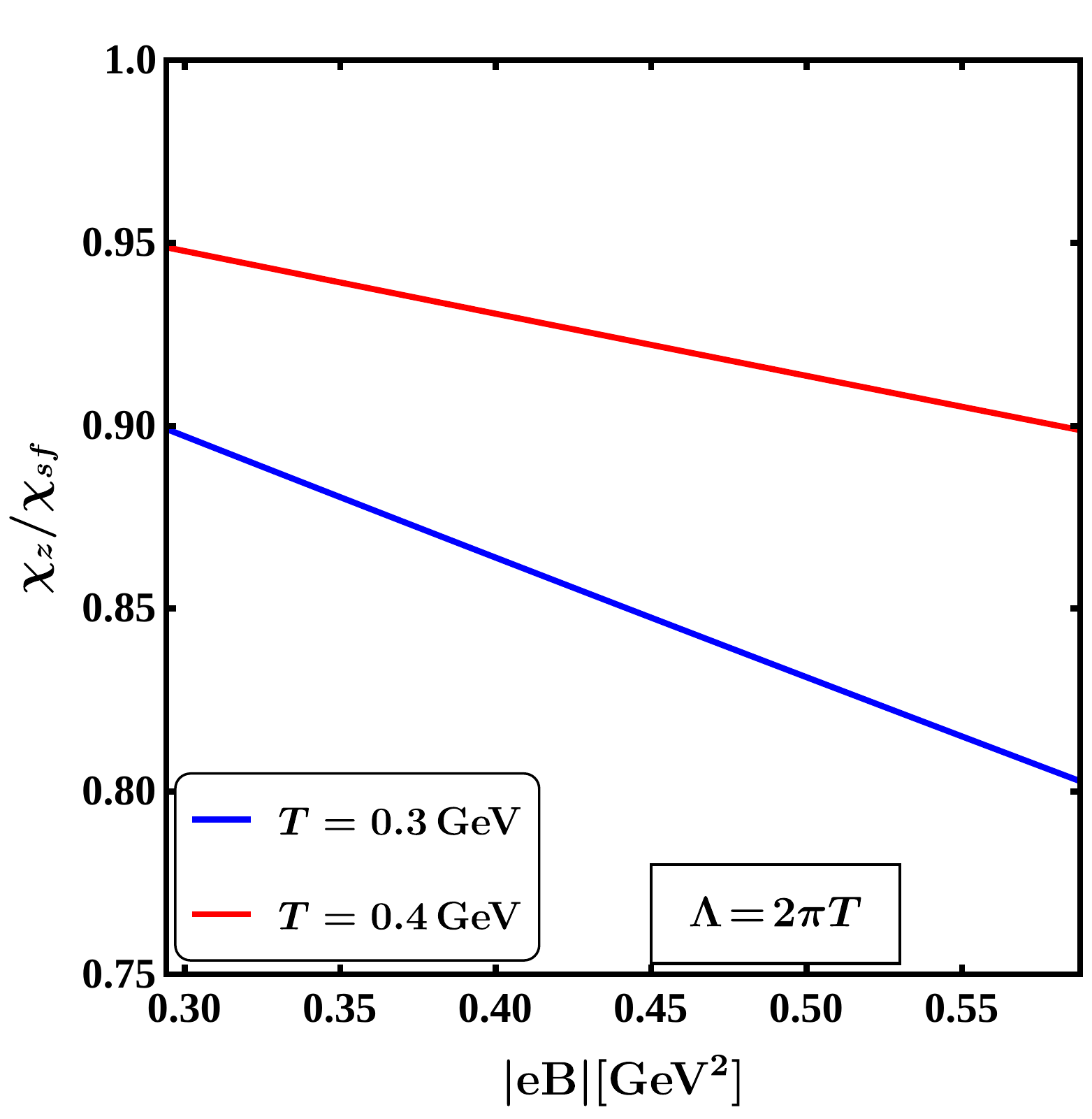} 
 \caption{Variation of the longitudinal part of the second-order QNS scaled with that of  free field value in presence of strong magnetic field with temperature (left panel) and magnetic field (right panel) strength for $N_f=3$.}
  \label{QNS_sfa_long_T}
 \end{center}
\end{figure}
\end{center} 
\begin{center}
\begin{figure}[tbh!]
 \begin{center}
 \includegraphics[scale=0.5]{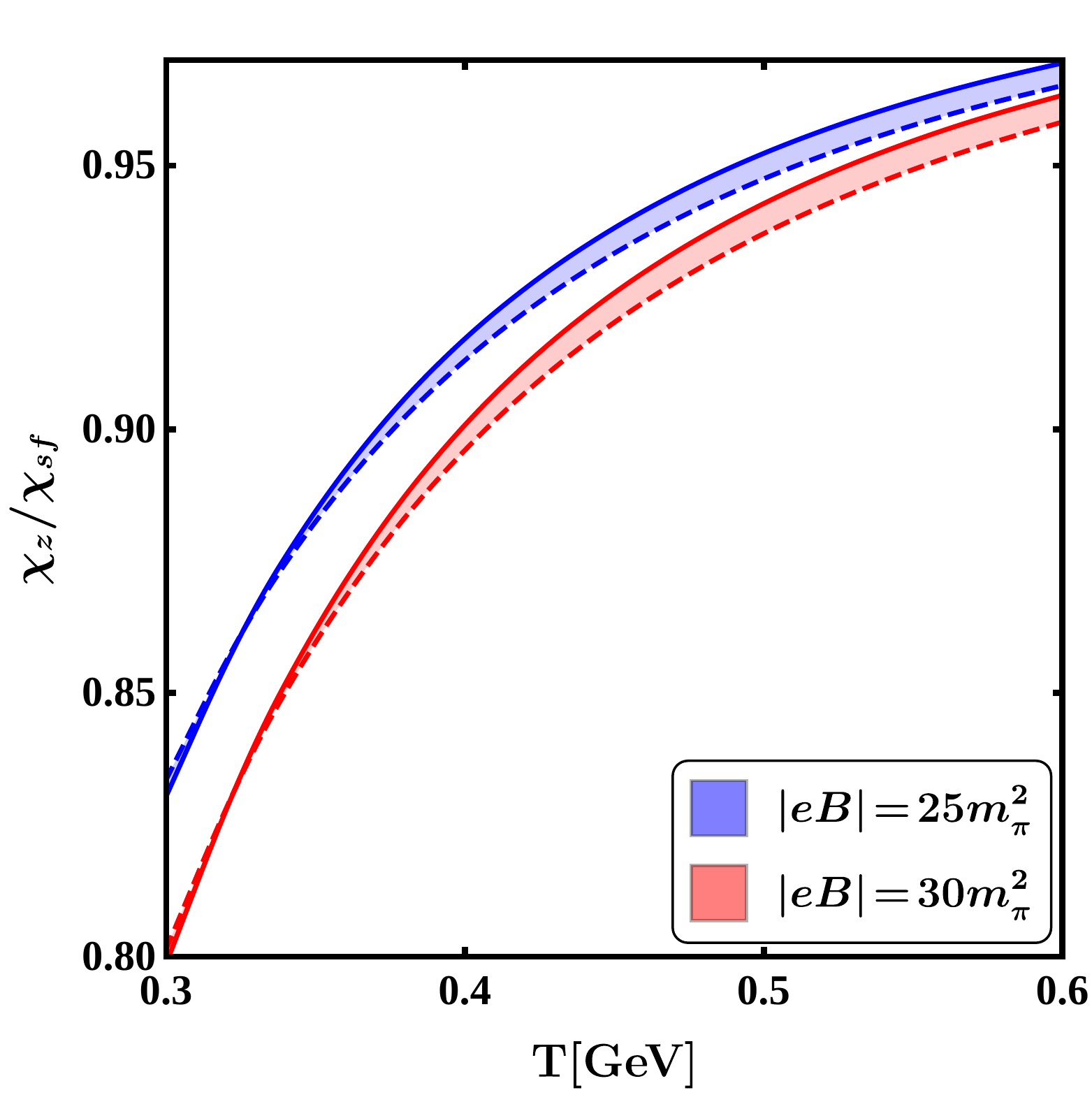} 
 \caption{Sensitivity of the longitudinal part of the second-order QNS scaled with that of free field value in presence of strong magnetic field on the renormalization scale for $N_f=3$. The dashed and the continuous curves represent $\Lambda=\pi T$ and $\Lambda=4\pi T$ respectively.}
  \label{QNS_sfa_long_lambda}
 \end{center}
\end{figure}
\end{center} 
In the left panel of Fig.~\ref{QNS_sfa_long_T} the variation of the longitudinal second-order QNS with temperature is displayed for two values of magnetic field strength and the central value of renormalization scale $\Lambda=2\pi T$. For a given magnetic field strength the longitudinal second-order QNS is found to increase with temperature and approaches the free field value at high temperature. On the other hand for a given temperature the longitudinal second-order QNS decreases with increase of the magnetic field strength as shown in the right panel of Fig. \ref{QNS_sfa_long_T} for two different temperatures and and the central value of renormalization scale $\Lambda=2\pi T$. 

The QGP pressure as well as the second order QNS is dependent on the renormalization scale $\Lambda$. Fig.~\ref{QNS_sfa_long_lambda} shows the sensitivity of the results on the choice of renormalization scale. Here we have varied it around the central value by a factor of two, i.e., from $\pi T$ to $4\pi T$.
%
\begin{center}
\begin{figure}[tbh!]
 \begin{center}
 \includegraphics[scale=0.5]{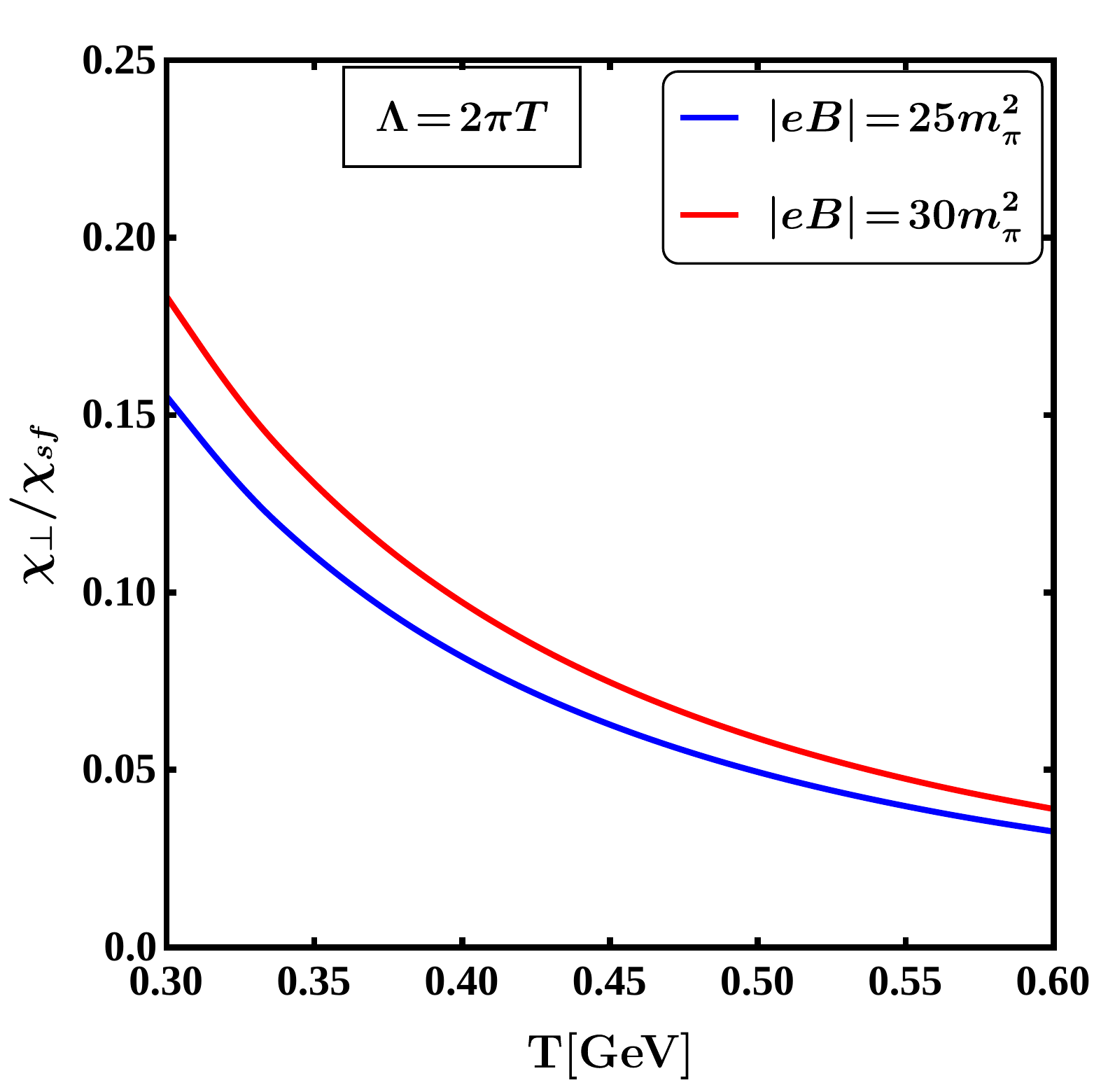} 
 \includegraphics[scale=0.5]{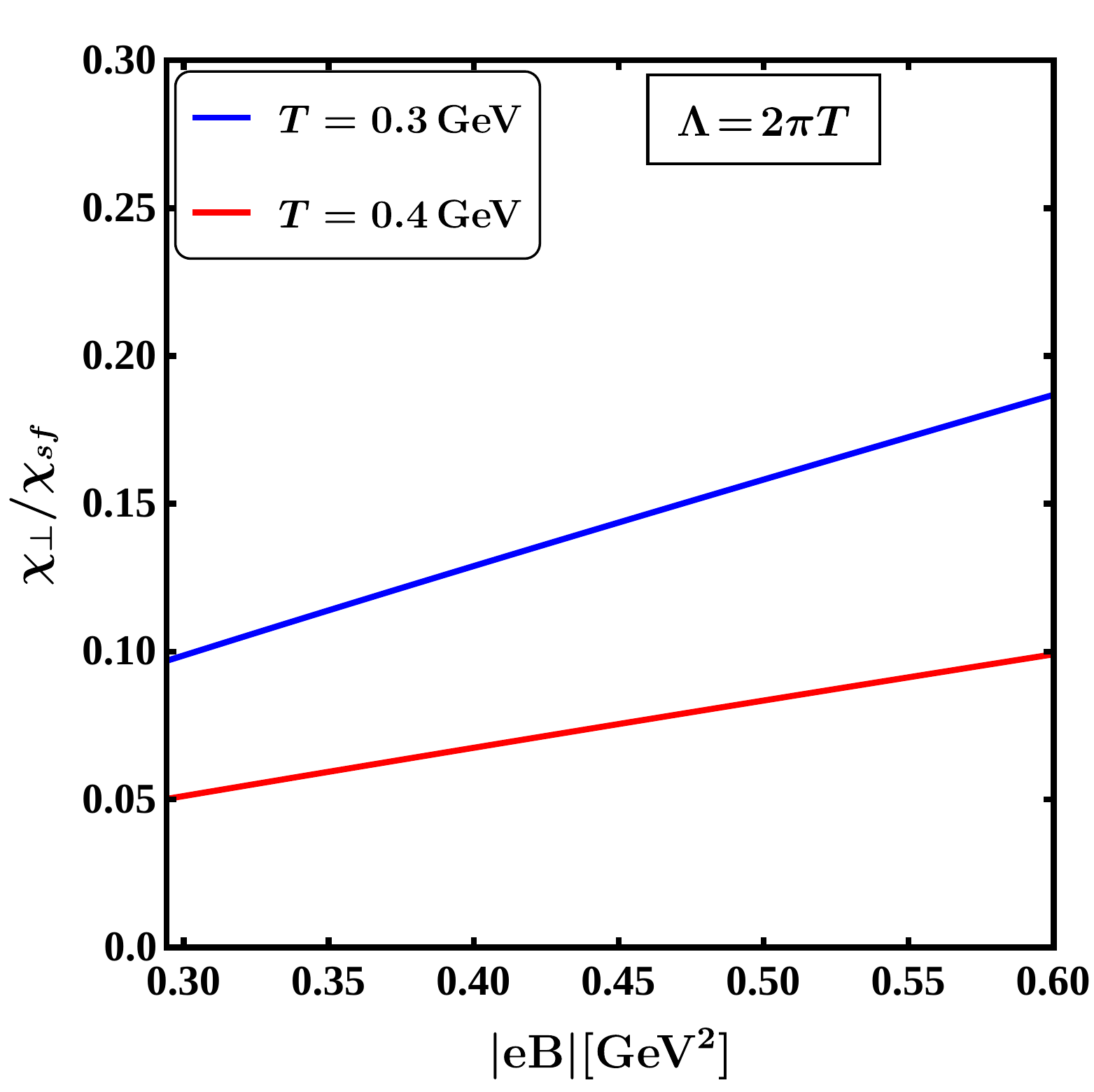} 
 \caption{Variation of the transverse part of the second-order QNS scaled with that of free field value in presence of strong magnetic field with temperature (left panel) and magnetic field (right panel) strength for $N_f=3$.}
  \label{QNS_sfa_trans_T}
 \end{center}
\end{figure}
\end{center} 
\begin{center}
\begin{figure}[tbh!]
 \begin{center}
 \includegraphics[scale=0.5]{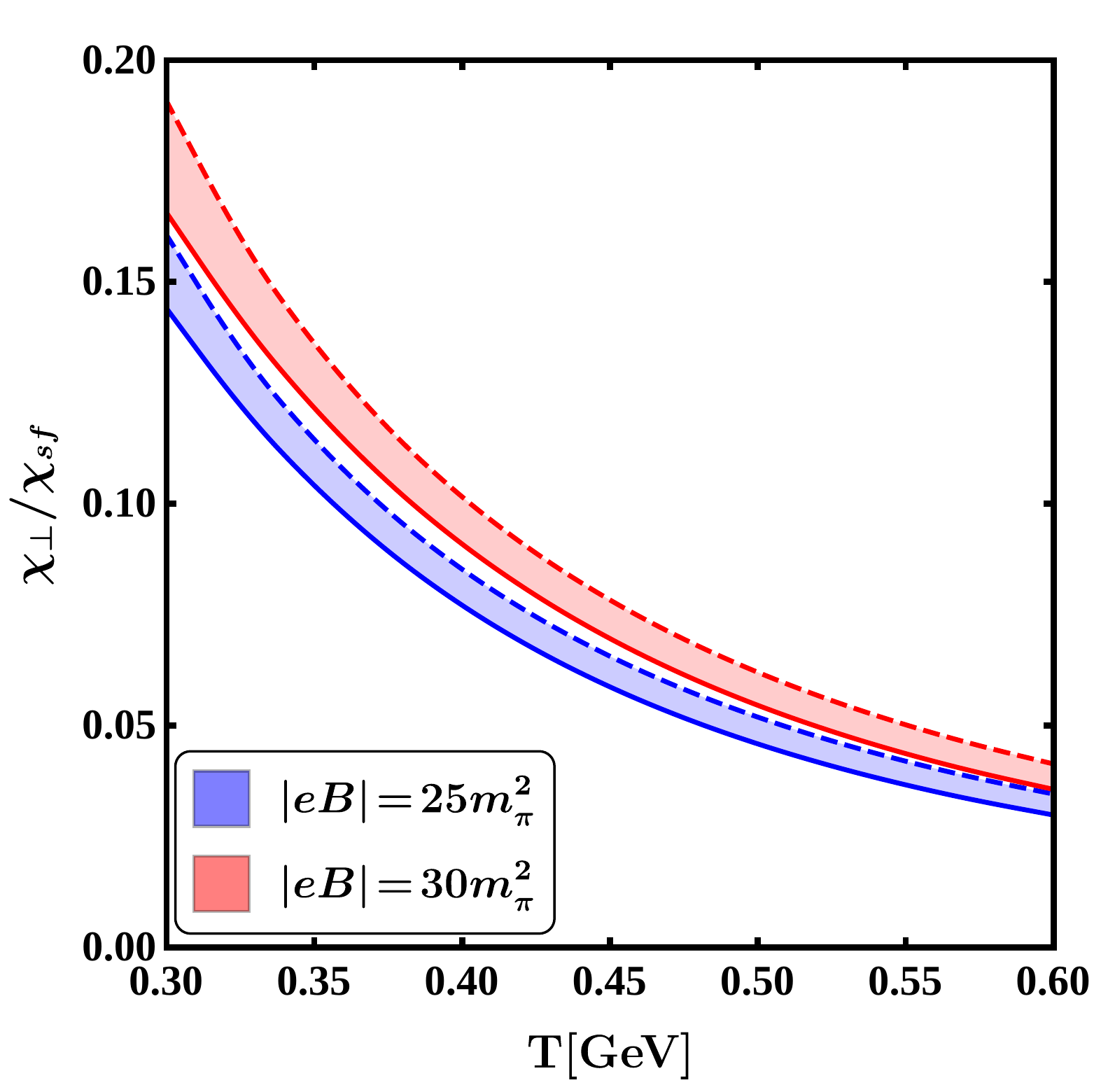} 
 \caption{Sensitivity of the transverse QNS scaled with that of free field value in presence of strong magnetic field on the renormalization scale for $N_f=3$. The dashed and the continuous curves represent $\Lambda=\pi T$ and $\Lambda=4\pi T$ respectively.}
  \label{QNS_sfa_trans_lambda}
 \end{center}
\end{figure}
\end{center} 
In the left panel of Fig.~\ref{QNS_sfa_trans_T} the variation of transverse second-order QNS with temperature is displayed for two values of magnetic field strength  and the central value of renormalization scale $\Lambda=2\pi T$. It is found that the
transverse second-order QNS decreases with temperature. This is an indication that the system may shrink in the transverse direction. For a given temperature the transverse second-order QNS is found to increase with the increase of the magnetic field strength as shown in the right panel of Fig.~\ref{QNS_sfa_trans_T} for two different temperatures  and the central value of renormalization scale $\Lambda=2\pi T$. This behaviour is in contrary to that of the longitudinal one. In Fig.~\ref{QNS_sfa_trans_lambda} the sensitivity of the transverse second order QNS on the renormalization scale is displayed by
varying a factor of two around the central value $\Lambda = 2\pi T$.

\begin{center}
\begin{figure}[tbh]
 \begin{center}
 \includegraphics[scale=0.5]{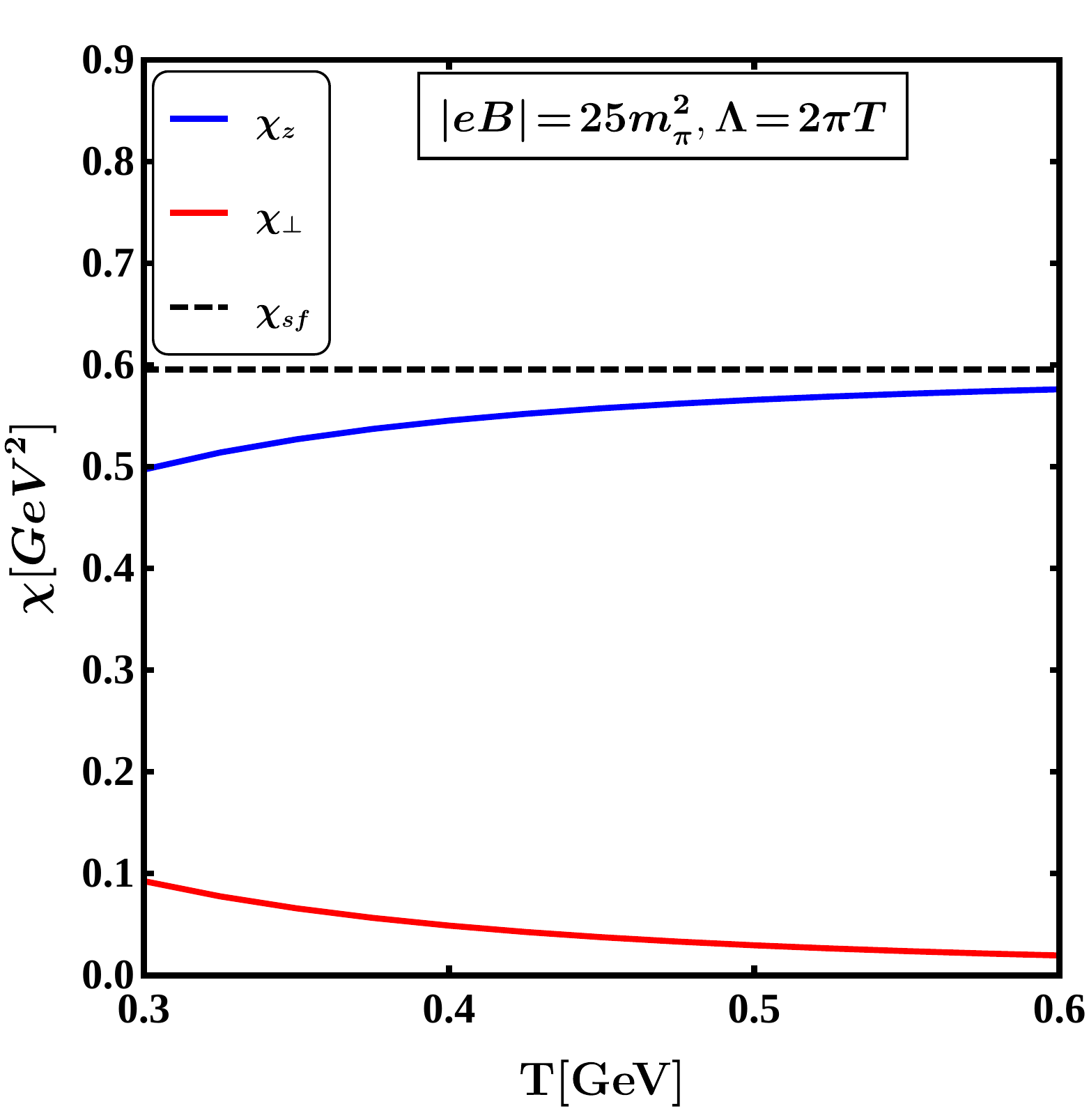}
 \caption{Variation of longitudinal and transverse QNS with temperature in presence of strong magnetic field.}
 \label{chi2_sfa_long_trans}
 \end{center}
\end{figure}
\end{center}

The second-order quark number susceptibility represents the fluctuation of net quark number over the average value. As the system becomes anisotropic in presence of strong magnetic field, we get two different pressure along the longitudinal and transverse direction to the magnetic field. It has been shown~\cite{Karmakar:2019tdp} in Fig.~\ref{P_mag_sfa} that the magnitude of the longitudinal pressure is greater than the transverse pressure. Thus the system expands more along the longitudinal direction. Similarly, one gets two different quark number susceptibility along the longitudinal and transverse direction. We can see from Eq.~\eqref{chi_sfa_ideal} that the longitudinal QNS of ideal quark gluon plasma in presence of strong magnetic field depends only on the strength of the magnetic field. However, the longitudinal QNS of the interacting quark gluon plasma depends both on temperature and magnetic field. This increases with temperature and matches with the ideal (non-interacting) QNS at very high temperatures which can be seen from Fig.~\ref{chi2_sfa_long_trans}.
However, transverse QNS behaves very differently from the longitudinal one due to the presence of magnetization which can be understood as follows. Transverse QNS of ideal quark gluon plasma is zero due to the fact that only gluon contributes to the ideal transverse pressure. Momentum of the quarks become restricted to the direction of magnetic field due to the dimensional reduction in presence of strong magnetic field. Hence the transverse pressure only consists of gluon pressure. Now, in presence of interaction one gets a non-zero transverse QNS because the transverse pressure gets contribution from internal quark loop. The transverse QNS gradually vanishes at high temperature (free limit) as can be seen from Fig.~\ref{chi2_sfa_long_trans}.

\section{Weak magnetic field}
\label{wfa}
In this section we consider magnetic field strength to be the lowest among all the scales $T$, $m_{th}$ as $\sqrt{|q_fB|} < m_{th}\sim gT <T$.  The HTL one-loop free-energy for the deconfined QCD matter has been calculated upto $\mathcal O[g^4]$ in Ref.~\cite{Bandyopadhyay:2017cle}. The total renormalized free-energy  in presence of weak magnetic field is sum of renormalized quark and gluon free-energy and can be written~\cite{Bandyopadhyay:2017cle}  as
\bea
F=F_q^r + F_g^r,
\eea
where the renormalized quark free-energy is
\bea
F_q^r &=& N_c N_f\Bigg[ 
-\frac{7\pi^2T^4}{180}\left(1+\frac{120\hat\mu^2}{7}+\frac{240\hat\mu^4}{7}
\right)+\frac{g^2C_FT^4}{48}\left(1+4\hat{\mu}^2\right)\left(1+12\hat{\mu}^2\right)\nn\\
&&+\, \frac{g^4C_F^2T^4}{
768\pi^2}\left(1+4\hat{\mu}^2\right)^2\left(\pi^2-6\right)+\frac{g^4C_F^2}{27N_f}
M_B^4 \bigg(12 \ln \frac{ \hat\Lambda}{2}-6\aleph(z)+\frac{36 \zeta (3)}{\pi^2}\nn\\
&&-2 -\frac{72}{\pi^2}\bigg)\Bigg].
\label{Eq:Fqr}
\eea
$M_{B,f}$ is the thermomagnetic mass for quark flavor $f$ in presence of weak magnetic field and $M_B$ represents flavor summed thermomagnetic quark mass as
\bea
M_B^2=\sum_f M_{B,f}^2 &=& \sum_f \frac{q_fB}{16\pi^2}\left[-\frac{1}{4}\aleph(z)-\frac{\pi 
T}{2m_f}-\frac{\gamma_E}{2}\right] . \label{mgmass}
\eea
$\aleph(z)$ in Eq.~\eqref{Eq:Fqr} is abbreviated as
\bea
\aleph(z) \equiv \Psi(z)+\Psi\left(z^{*}\right),
\eea
with $\Psi(z)$ is the digamma function
\bea
\Psi(z) \equiv \frac{\Gamma^{\prime}(z)}{\Gamma(z)},
\eea
and $z=1/2 -i\hat{\mu}$. 
At small chemical potential, $\aleph(z)$ can be expanded as
\bea
 \aleph(z)&=&-2\gamma_E-4\ln 
2+14\zeta(3)\hat{\mu}^2-62\zeta(5)\hat{\mu}^4+254\zeta(7)\hat{\mu}^6+{\cal 
O}(\hat{\mu}^8).\label{aleph}
\eea
In addition to the renormalized quark free-energy in Eq.~\eqref{Eq:Fqr}, the renormalized gluon free-energy is given as
\bea
\hspace{-1cm}\frac{F_g^r}{d_A}&=&-\frac{\pi^2 T^4}{45}\left[1-\frac{15}{2}\hat{m}_D^2+30(\hat{m}_D^w)^3+\frac{45}{8}\hat{m}_D^4\(2\ln\frac{\hat{\Lambda}}{2}-7+2\gamma_E+\frac{2\pi^2}{3}\)\right]\nn\\
&-&\pi^2T^4\hat{m}_D^2\delta \hat{m}_D^2\left(\gamma_E+\ln\hat\Lambda\right)+\sum_f\frac{g^2(q_fB)^2}{(12\pi)^2}\frac{T^2}{m_f^2}\Bigg[4.97+2\ln\frac{\hat\Lambda}{2}\nn\\
&+&3\hat{m}_D^2\Bigg\{2\(1-\ln 2\)\ln^2\frac{\hat\Lambda}{2}+2\bigg(\frac{7}{2}-\frac{\pi ^2}{6}-\ln ^2(2)-2 \gamma_E  (\ln 2-1)\bigg)\ln\frac{\hat\Lambda}{2}+4.73\Bigg\}\Bigg]\nn\\
&-& \sum_f\frac{g^2(q_fB)^2}{(12\pi)^2}\frac{\pi T}{32m_f}\Biggl[\Biggl\{\frac{3}{4}\ln^2\frac{\hat\Lambda}{2}+2\ln\frac{\hat\Lambda}{2}\(\frac{21}{8}+\frac{3}{4}\frac{\zeta'(-1)}{\zeta(-1)}+\frac{27}{4}\ln 2\)+43.566\nn\\
&+&\frac{3}{4}\hat{m}_D^2\Bigg[2\ln^2\frac{\hat\Lambda}{2}\(5\pi ^2-\frac{609}{10}+\frac{114 \ln 2}{5}\) + 2\ln\frac{\hat\Lambda}{2}\bigg( 30 \zeta (3)-\frac{5779}{75}+\frac{121}{6} \pi ^2+\frac{114}{5}\ln ^2(2)\nn\\
&+&\frac{468}{25} \ln 2+ \gamma_E  \(10\pi ^2-\frac{609}{5}+\frac{228}{5} \ln 2\)\bigg)+106.477\Bigg]\Bigg\}+\frac{8}{3\pi}\Bigg\{\(3\ln 2-4\)\ln\frac{\hat\Lambda}{2}-3.92\nn\\
&+&3\hat{m}_D^2\Bigg[\frac{1}{20}\ln^2\frac{\hat\Lambda}{2}\Big(11+5\pi ^2-92 \ln 2\Big)+
2\ln\frac{\hat\Lambda}{2}\(\frac{3}{4}\zeta (3)+\frac{1557}{200}- \frac{\pi ^2}{3}-\frac{23}{10}\ln ^2(2)\right.\nn\\
&-&\left.\frac{168}{25} \ln 2+\gamma_E\(\frac{11}{20}+\frac{\pi^2}{4}-\frac{23}{5}\ln 2\)\)-1.86\Bigg]\Bigg\}\Biggr],
\label{Eq:Fgr}
\eea
where $\hat m_D^w=m_D^w/2\pi T$, $\hat m_D=m_D/2\pi T$, $ \delta \hat m_D=\delta m_D/2\pi T$ and $m_D^w$ represents the Debye mass in weak magnetic field approximation and is obtained as
\bea
\(m_D^w\)^2 &\simeq& 
\frac{g^2 T^2}{3}\left[\left(N_c+\frac{N_f}{2}\right)+6N_f\hat{\mu}
^2\right] \nn\\
&+&\sum\limits_f \frac{g^2(q_fB)^2}{12\pi^2T^2} 
\sum\limits_{l=1}^\infty 
(-1)^{l+1}l^2 \cosh\left(2l\pi\hat{\mu}\right) K_0\left(\frac{m_fl}{T}\right) + 
\mathcal{O}[(q_fB)^4] \nn \\
&=& m_D^2 + \delta m_D^2.
\label{md_wfa}
\eea
Considering the expression of free energy vis-a-vis  pressure we calculate the second-order QNS in weak field limit by using Eq.~\eqref{chi_def}.
The second-order QNS of free quarks and gluons in thermal medium is given as
\bea
\chi_f=\frac{1}{3}N_c N_f T^2.
\eea
\begin{center}
\begin{figure}[tbh]
 \begin{center}
 \includegraphics[scale=0.57]{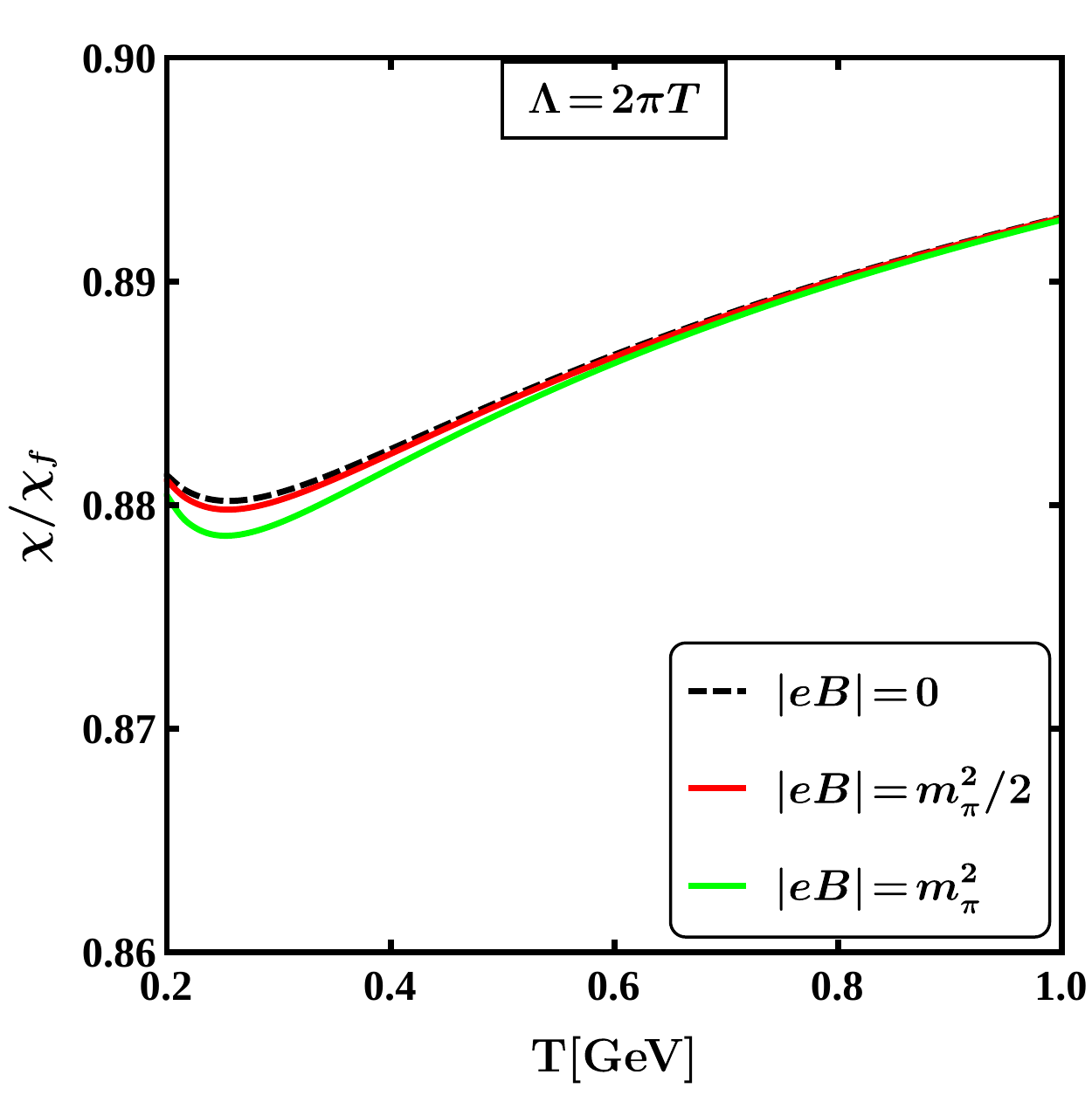} 
 \includegraphics[scale=0.6]{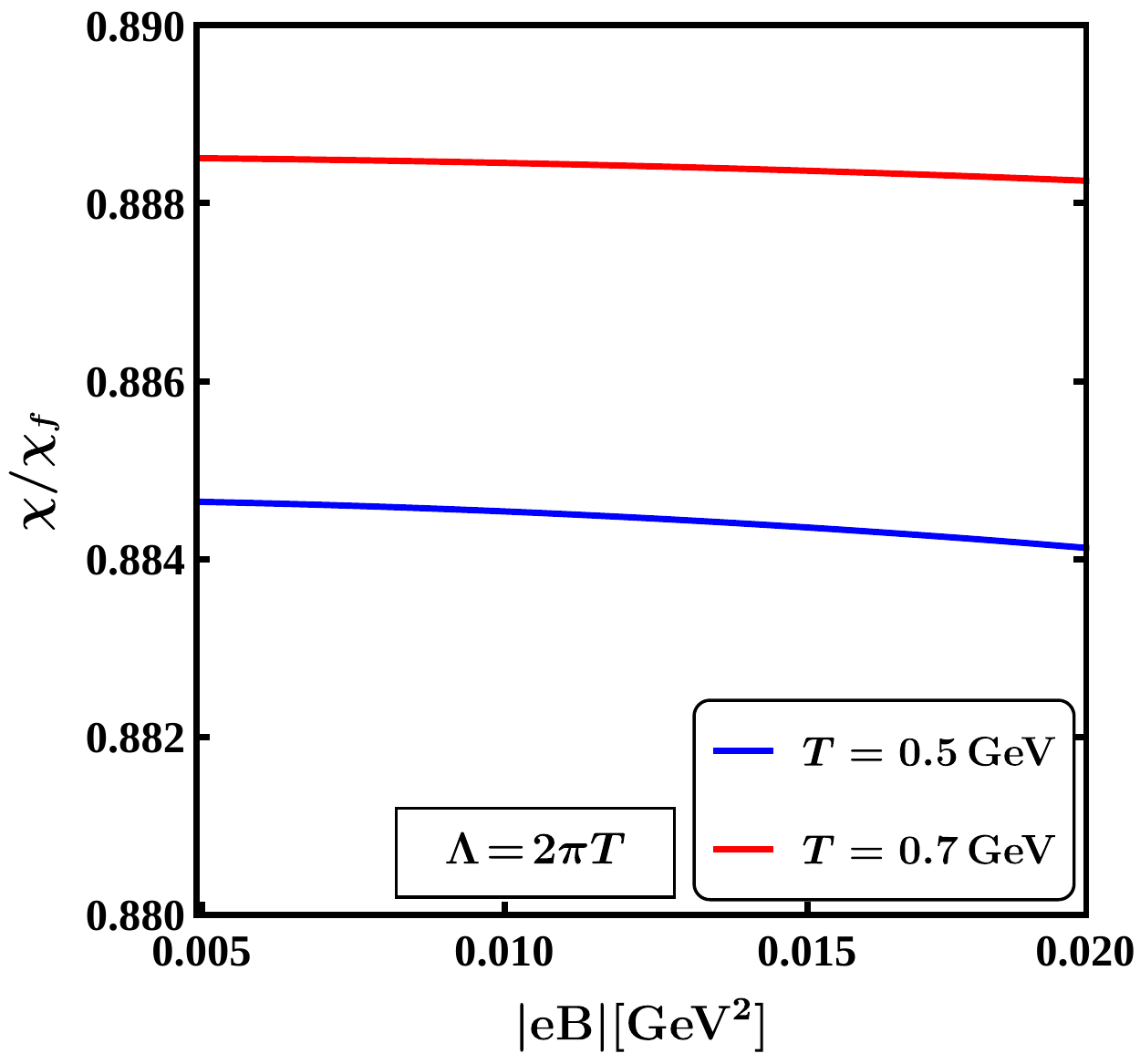} 
 \caption{Variation of second-order QNS  scaled with thermal free field value with temperature (left panel) and magnetic field strength (right panel) for $m_f=5$ MeV and $N_f=3$.}
  \label{QNS_wfa}
 \end{center}
\end{figure}
\end{center} 
\begin{center}
\begin{figure}[tbh!]
 \begin{center}
 \includegraphics[scale=0.6]{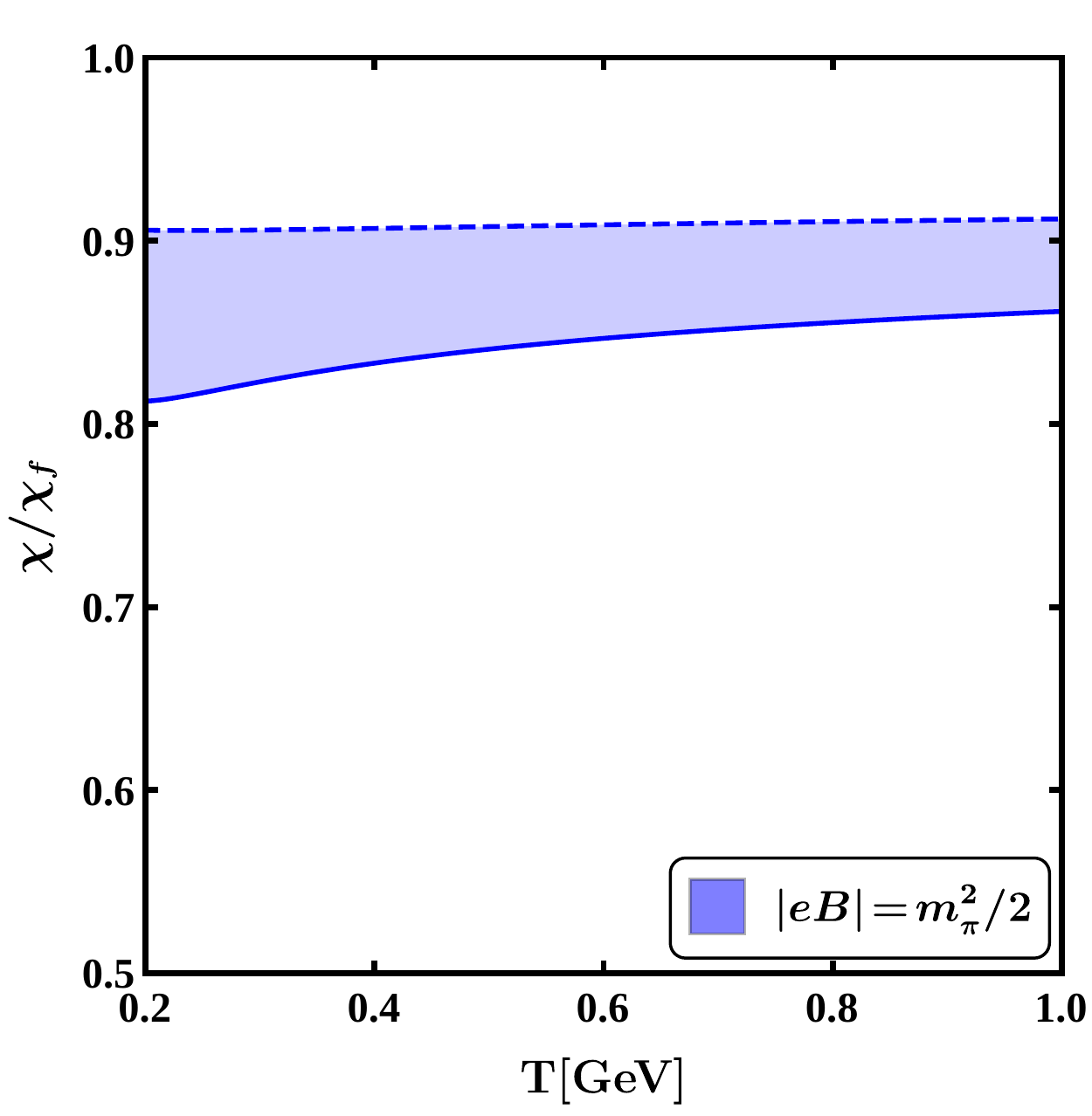} 
 \caption{Sensitivity of the second-order QNS scaled with that of free field value in presence of weak magnetic field on the renormalization scale for $N_f=3$. The dashed and the continuous curves represent $\Lambda=\pi T$ and $\Lambda=4\pi T$ respectively.}
  \label{QNS_wfa_lambda}
 \end{center}
\end{figure}
\end{center} 
The left panel of Fig.~\ref{QNS_wfa} shows the variation of the scaled second-order QNS with the temperature at different values of the magnetic field strength and the central value of renormalization scale $\Lambda=2\pi T$. The weak field effect appears as a correction to the thermal medium, the weak field second-order QNS is not very much different than that of thermal medium. It is found to increase with temperature and approaches the free field value at high enough temperature. The magnetic field effect on the second-order QNS is visible at low temperature. The value of second-order QNS slowly decreases as one increases the magnetic field strength as shown in the right panel of Fig.~\ref{QNS_wfa}. In Fig.~\ref{QNS_wfa_lambda} the sensitivity of the second order weak field QNS on the renormalization scale is displayed by
varying a factor of two around the central value $\Lambda = 2\pi T$.
\section{Conclusion}
\label{conclusion}
We consider a hot and dense deconfined QCD matter in the presence of the background strong and weak magnetic field within HTL approximation. The quarks are directly affected by magnetic field whereas gluons are affected via quark loop in the gluon self-energy. In the strong field approximation we assume quarks are in lowest Landau level. We compute the one-loop HTL pressure in the presence of finite temperature and chemical potential in the lowest Landau level within the strong field approximation. Various divergent terms are eliminated by choosing appropriate counterterms in the ${\overline {\mbox{MS}}}$ renormalization scheme. The presence of magnetization causes the system to be anisotropic, and one obtains two different pressures in directions parallel and perpendicular to the magnetic field. Both the longitudinal  and transverse  pressures are computed analytically by calculating the magnetization of the system. We then compute both the longitudinal and transverse second-order QNS in the strong field approximation. For a given magnetic field strength, the longitudinal second-order QNS increases with temperature and approaches the non-interacting value at high enough temperature. For a given temperature the longitudinal second-order QNS is found to decrease with increase of magnetic field strength. In contrast the transverse second-order QNS is found to decrease with temperature and increase with the increase of magnetic field. Further, in weak field approximation we consider  one-loop HTL pressure of hot and dense QCD matter of Ref.~\cite{Bandyopadhyay:2017cle} and compute the second-order QNS. The thermomagnetic correction is found to be marginal and slowly varies with magnetic field. Our calculation can be compared with future lattice QCD calculation.

\section{ACKNOWLEDGMENTS}
BK and MGM were funded by Department of Atomic Energy (DAE), India via the project TPAES. NH was funded by DAE, India. BK acknowledges useful discussions with Arghya Mukherjee and Aritra Bandyopadhyay.
\appendix
\section{Calculation of the quark self-energy form factors}
\label{quark_ff}
\subsection{Calculation of the form factors $a$ and $d$}

We denotes the momentum four-vectors as $K^{\mu}=(k_0,k_1,k_2,k_3)$ and we decompose the four-vector $K^{\mu}$ into its parallel and perpendicular components as $K_\sp^\mu=(K \cdot u)u^\mu-(K \cdot n)n^\mu=(k_0,0,0,k_3)$ and $K_\perp^\mu=K^\mu-K_\sp^\mu=(0,k_1,k_2,0)$. Thus the scalar product becomes $K_\perp^\mu\cdot (K_\perp)_\mu= K_\perp^2=-k_\perp^2=-k_1^2-k_2^2$.

Now, we can calculate the form factor $a$ from Eq.~\eqref{a_def} as
\bea
a=\frac{1}{4}\Tr[\Sigma\, \slashed u\, ]&=&- 2g^2 C_F \sumintf_{\{k_0\}} e^{-\frac{k_\perp^2}{q_fB}}\left[\frac{k_0}{K_\sp^2(K-P)^2_\sp} + (k-p)_\perp^2\frac{k_0}{K_\sp^2 (K-P)^4_\sp}\right]\label{a_sumint}
\eea
where $g$ is the QCD coupling constant, $C_F=4/3$ is the Casimir color-factor associated  with gluon emission from a quark and $q_f$ is the charge of the fermion of flavor $f$. The sum-integral is given as
\bea
\sumintf_{\{k_0\}}\equiv T\!\!\!\!\! \sum_{k_0=(2n+1)\pi T i + \mu 
		}  \int \frac{d^3k}{(2\pi)^3}. 
\eea

So, Eq.~\eqref{a_sumint} becomes
\bea
a&=& -2g^2 C_F  \int \frac{d^3k}{(2\pi)^3} e^{-\frac{k_\perp^2}{q_fB}}\left[T_2 + (k-p)_\perp^2 T_4\right]\nn\\
&=& -2g^2 C_F  \int_{-\infty}^{\infty} \frac{dk_3}{2\pi} \left[ \frac{q_fB}{4\pi} T_2 + \frac{q_f B}{4\pi}(p_{\perp}^2+q_fB)\hspace{.2cm} T_4\right]\nn\\
&=&- \frac{g^2 C_F (q_f B)}{4\pi^2}  \int_{-\infty}^{\infty} dk_3 \left[  T_2 + (p_{\perp}^2+q_fB)\hspace{.2cm} T_4\right],\label{a}
\eea
where
\bea
T_2&=& \sum_{\{k_0\}} \frac{k_0}{K_\sp^2(K-P)^2_\sp},\nn\\
T_4&=&  \sum_{\{k_0\}} \frac{k_0}{K_\sp^2 (K-P)^4_\sp}
= -\frac{1}{2k_3}\frac{\partial T_2}{\partial p_3}.
\eea
Here we also note that in LLL, $p_{\perp}=0$. Now we perform the Matsubara sum~\cite{Bellac:2011kqa} and use HTL approximations (loop momentum $\sim T$, external momentum $\sim gT$).
\bea
T_2&=& \sum_{\{k_0\}} \frac{k_0}{K_\sp^2(K-P)^2_\sp},\nn\\
&=&-\frac{1}{4k_3 }\Bigg[\frac{n_B(k_3)+n_F(k_3-\mu)}{p_0+p_3}+\frac{n_B(k_3)+n_F(k_3+\mu)}{p_0-p_3}\Bigg]
\eea
We use the following equations to perform the sum-integrals.
\bea
\int_{-\infty}^{\infty} \frac{dk_3}{k_3} \,n_F(k_3\pm \mu)&=&2\int_{0}^{\infty} \frac{dk_3}{k_3}\,n_F(k_3\pm \mu),\nn\\
\int_{-\infty}^{\infty} \frac{dk_3}{k_3^2} \,n_F(k_3\pm \mu)&=&2\int_{0}^{\infty} \frac{dk_3}{k_3^2}\, n_F(k_3\pm  \mu).
\eea
Hence,
\bea
&&\int_{-\infty}^{\infty} dk_3 \hspace{.2cm} T_2\nn\\
&=&- \int_0^{\infty}\frac{dk_3}{4k_3}\bigg[ \frac{2n_B(k_3)+2n_F(k_3-\mu)}{p_0+p_3} + \frac{2n_B(k_3)+2n_F(k_3+\mu)}{p_0-p_3} \bigg]\nn\\
&=&-\int_0^{\infty}\frac{dk_3}{4k_3}\bigg( 2n_B(k_3) \frac{2p_0}{p_0^2-p_3^2}   +\frac{2n_F(k_3+\mu)}{p_0-p_3}+\frac{2n_F(k_3-\mu)}{p_0+p_3}\bigg)\nn\\
&=&\frac{p_0}{p_0^2-p_3^2}\bigg[\ln 2-\frac{\mu^2}{T^2}\frac{7\zeta(3)}{8\pi^2}+\frac{\mu^4}{T^4} \frac{31\zeta(5)}{32\pi^4} \bigg] +\frac{p_3}{p_0^2-p_3^2}\bigg[\frac{\mu }{4T}\Big(-1 -\gamma_E-\frac{4}{3}\ln 2 \nn\\
&+&12\ln G\Big)+\frac{\mu^3}{48T^3}\Big( \gamma_E+\frac{16}{15}\ln 2-120\zeta'(-3)\Big)+\mathcal O[\left(\mu/T\right)^5\bigg],
\eea
where $G\approx1.2824$ is Glaisher's constant and $\zeta'(z)=\frac{d\zeta(z)}{dz}=-\sum_{n=2}^{\infty}\frac{\ln n}{n^z}$.
\bea
&&\int_{-\infty}^{\infty} dk_3 \,q_fB \,T_4\nn\\
&=& -\frac{q_fB}{2}\frac{\partial}{\partial p_3}\int_{-\infty}^{\infty} \frac{dk_3}{k_3}\,T_2\nn\\
&=& \frac{q_fB}{2}\frac{\partial}{\partial p_3}\int_{-\infty}^{\infty} \frac{dk_3}{4k_3^2}\bigg(\frac{n_B(k_3)+n_F(k_3-\mu)}{p_0+p_3}+\frac{n_B(k_3)+n_F(k_3+\mu)}{p_0-p_3}\bigg)\nn\\
&=& \frac{q_fB}{2}\frac{\partial}{\partial p_3}\int_{0}^{\infty} \frac{dk_3}{4k_3^2}\bigg( 2n_B(k_3) \frac{2p_0}{p_0^2-p_3^2}   +\frac{2n_F(k_3+\mu)}{p_0-p_3}+\frac{2n_F(k_3-\mu)}{p_0+p_3}\bigg)\nn\\
&=&\frac{q_fB}{T}\Bigg[\frac{p_0 p_3}{(p_0^2-p_3^2)^2}\bigg\{\frac{1}{6}\left(\gamma_E+2\ln 2 -12 \ln G\right)+\frac{\mu^2}{16T^2}\Big( 1-\gamma_E-\frac{16}{15}\ln 2\nn\\
&+&120\zeta'(-3) \Big) +\frac{\mu^4}{96T^4}\Big(  -1+\gamma_E+\frac{64}{63}\ln 2+252\zeta'(-5)\Big)\bigg\}+\frac{p_0^2+p_3^2}{(p_0^2-p_3^2)^2}\nn\\
&\times&\bigg\{\frac{\mu}{T}\frac{7\zeta(3)}{8\pi^2}-\frac{\mu^3}{T^3}\frac{31\zeta(5)}{16\pi^4}  \bigg\}+\mathcal O[\left(\mu/T\right)^5\Bigg].
\eea
So the form factor $a=-d$ upto $\mathcal O[(\mu/T)^4]$ can be written in compact form as
\bea
a=-d
&=&c_1 \Bigg[\frac{p_0}{P_\sp^2}c_2 +\frac{p_3}{P_\sp^2}c_3+\frac{p_0p_3}{P_\sp^4}c_4 + \Big(\frac{1}{P_\sp^2}+\frac{2p_3^2}{P_\sp^4} \Big)c_5  \Bigg]\label{ad_final}
\label{ad_final}
\eea
where
\bea
c_1&=&- \frac{g^2 C_F (q_f B)}{4\pi^2},\nn\\
c_2
&=&\bigg[\ln 2-\frac{\mu^2}{T^2} \frac{7\zeta(3)}{8\pi^2} +\frac{\mu^4}{T^4}\frac{31\zeta(5)}{32\pi^4} \bigg],\nn\\
c_3
&=&\bigg[\frac{\mu }{4T}\Big( -1-\gamma_E-\frac{4}{3}\ln 2+12\ln G\Big)+\frac{\mu^3}{48T^3}\Big( \gamma_E+\frac{16}{15}\ln 2-120\zeta'(-3)\Big)\bigg],\nn\\
c_4
&=&\frac{q_fB}{T}\bigg[\frac{1}{6}\left(\gamma_E+2\ln 2 -12 \ln G\right)+\frac{\mu^2}{16T^2}\Big( 1-\gamma_E-\frac{16}{15}\ln 2+120\zeta'(-3)\Big)\nn\\
&+&\frac{\mu^4}{96T^4}\Big(  -1+\gamma_E+\frac{64}{63}\ln 2+252\zeta'(-5)\Big)\bigg],\nn\\
c_5
&=&\frac{q_fB}{T}\bigg[\frac{\mu}{T}\frac{7\zeta(3)}{8\pi^2}-\frac{\mu^3}{T^3}\frac{31\zeta(5)}{16\pi^4} \bigg]\label{c1c2c3}.
\eea
\subsection{Calculation of quark form factor $b$ and $c$}
Similarly one can calculate $b$ from Eq.~\eqref{b_def} as
\bea
b=-\frac{1}{4}\Tr[\Sigma\, \slashed n\, ]&=& 2g^2 C_F \sumintf_{\{k_0\}} e^{-\frac{k_\perp^2}{q_fB}}\left[\frac{k_3}{K_\sp^2(K-P)^2_\sp} + (k-p)_\perp^2\frac{k_3}{K_\sp^2 (K-P)^4_\sp}\right]\nn\\
&=& 2g^2 C_F  \int \frac{d^3k}{(2\pi)^3} e^{-\frac{k_\perp^2}{q_fB}}\,k_3\,\left[T_1 + (k-p)_\perp^2 T_3\right]\nn\\
&=& 2g^2 C_F  \int_{-\infty}^{\infty} \frac{dk_3}{2\pi} \,k_3\,\left[ \frac{q_fB}{4\pi} T_1 + \frac{q_f B}{4\pi}(q_fB)\hspace{.2cm} T_3\right]\nn\\
&=& \frac{g^2 C_F (q_f B)}{4 \pi^2}  \int_{-\infty}^{\infty} dk_3 \hspace{.2cm}k_3 \left[  T_1 +q_fB\hspace{.2cm} T_3\right],
\label{b}
\eea
where
\bea
T_1&=& \sum_{\{k_0\}} \frac{1}{K_\sp^2(K-P)^2_\sp},\label{T1_def}\\
T_3&=& \sum_{\{k_0\}} \frac{1}{K_\sp^2 (K-P)^4_\sp}
=- \frac{1}{2k_3}\frac{\partial T_1}{\partial p_3}.
\eea
After doing the Matsubara sum, Eq.~\eqref{T1_def} becomes

\bea
T_1&=& \frac{1}{4k_3^2 }\Bigg[\frac{n_B(k_3)+n_F(k_3-\mu)}{p_0+p_3}-\frac{n_B(k_3)+n_F(k_3+\mu)}{p_0-p_3}\Bigg].\nn\\
\eea
Hence,
\bea
&&\int_{-\infty}^{\infty} dk_3 \hspace{.2cm} k_3\, T_1\nn\\
&=& \int_0^{\infty}\frac{dk_3}{4k_3}\bigg[ \frac{2n_B(k_3)+2n_F(k_3-\mu)}{p_0+p_3}-\frac{2n_B(k_3)+2n_F(k_3+\mu)}{p_0-p_3} \bigg]\nn\\
&=&-\int_0^{\infty}\frac{dk_3}{4k_3}\bigg( 2n_B(k_3)\frac{2p_3}{p_0^2-p_3^2}+\frac{2n_F(k_3+\mu)}{p_0-p_3}-\frac{2n_F(k_3-\mu)}{p_0+p_3}\bigg) \nn\\
&=&\frac{p_3}{p_0^2-p_3^2}\bigg[\ln 2-\frac{\mu^2}{T^2}\frac{7\zeta(3)}{8\pi^2}+\frac{\mu^4}{T^4}  \frac{31\zeta(5)}{32\pi^4}\bigg]+\frac{p_0}{p_0^2-p_3^2}\bigg[\frac{\mu}{4T}\Big( -1 -\gamma_E-\frac{4}{3}\ln 2\nn\\
&+&12 \ln G\Big) +\frac{\mu^3}{48T^3}\Big( \gamma_E+\frac{16}{15}\ln 2-120\zeta'(-3) \Big)+\mathcal O[(\mu/T)^5\bigg]
\eea
and
\bea
&&\int_{-\infty}^{\infty} dk_3 \, k_3\,q_fB \,T_3\nn\\
&=& -\frac{q_fB}{2}\frac{\partial}{\partial p_3}\int_{-\infty}^{\infty} dk_3\,T_1\nn\\
&=& -\frac{q_fB}{2}\frac{\partial}{\partial p_3}\int_{-\infty}^{\infty} \frac{dk_3}{4k_3^2}\Bigg[\frac{n_B(k_3)+n_F(k_3-\mu)}{p_0+p_3}-\frac{n_B(k_3)+n_F(k_3+\mu)}{p_0-p_3}\Bigg]\nn\\
&=& \frac{q_fB}{2}\frac{\partial}{\partial p_3}\int_{0}^{\infty} \frac{dk_3}{4k_3^2}\Bigg[\frac{2n_B(k_3)+2n_F(k_3-\mu)}{p_0+p_3}-\frac{2n_B(k_3)+2n_F(k_3+\mu)}{p_0-p_3}\Bigg]\nn\\
&=&- \frac{q_fB}{2}\frac{\partial}{\partial p_3}\int_{0}^{\infty} \frac{dk_3}{4k_3^2}\Bigg[2n_B(k_3)\frac{2p_3}{p_0^2-p_3^2}+\frac{2n_F(k_3+\mu)}{p_0-p_3}-\frac{2n_F(k_3-\mu)}{p_0+p_3}\Bigg]\nn\\
&=&-\frac{q_fB}{T}\Bigg[\frac{p_0^2+ p_3^2}{(p_0^2-p_3^2)^2}\bigg\{\frac{1}{6}\left(\gamma_E+2\ln 2 -12 \ln G\right)+\frac{\mu^2}{16T^2}\Big( 1-\gamma_E-\frac{16}{15}\ln 2\nn\\
&+&120\zeta'(-3) \Big) +\frac{\mu^4}{96T^4}\Big(  -1+\gamma_E+\frac{64}{63}\ln 2+252\zeta'(-5)\Big)\bigg\}+\frac{4p_0 p_3}{(p_0^2-p_3^2)^2}\nn\\
&\times&\bigg\{\frac{\mu}{T}\frac{7\zeta(3)}{8\pi^2}-\frac{\mu^3}{T^3} \frac{31\zeta(5)}{16\pi^4}  \bigg\}+\mathcal O[\left(\mu/T\right)^5 \Bigg].
\eea
The form factor $b=-c$ is obtained upto $\mathcal O[(\mu/T)^4]$ as 
\bea
b=-c
&=&-c_1 \Bigg[\frac{p_3}{P_\sp^2}c_2 +\frac{p_0}{P_\sp^2}c_3-\Big(\frac{1}{P_\sp^2}+\frac{2p_3^2}{P_\sp^4} \Big)c_4 - \frac{4p_0p_3}{P_\sp^4}c_5 \Bigg].
\label{bc_final}
\eea
\section{One-loop sum-integrals for quark free-energy}
\label{quark_free_energy}
Eq.~\eqref{Fq'_ini} can be rewritten as
\bea
F'_q&=& -4 d_F \sum_f  \frac{q_fB}{(2\pi)^2}\sumintf_{\{p_0\}}dp_3~\bigg[\frac{a p_{0}}{P_{\sp}^{2}}+\frac{b p_{3}}{P_{\sp}^{2}}
-\frac{a^{2}}{P_{\sp}^{2}}-\frac{b^{2}}{P_{\sp}^{2}}-\frac{2 a^{2} p_{3}^{2}}{P_{\sp}^{4}}-\frac{2 b^{2} p_{3}^{2}}{P_{\sp}^{4}}-\frac{4 a b p_{0} p_{3}}{P_{\sp}^{4}}\bigg]
\label{Fq'_2}
\eea
The various sum-integrals in Eq.~\ref{Fq'_2} can be written using Eq.~\eqref{ad_final} and Eq.~\eqref{bc_final} as
\bea
\sumintf_{\{p_0\}} ~\frac{ap_0}{P_{\sp}^2}&=& c_1 \sumintf_{\{p_0\}} \bigg[c_2\bigg(\frac{1}{P_{\sp}^2}+\frac{p_3^2}{P_\sp^4}\bigg)+\frac{p_0p_3}{P_\sp^4}c_3+\bigg( \frac{p_3}{P_\sp^4}+\frac{p_3^3}{P_\sp^6} \bigg)c_4+c_5\(\frac{p_0}{P_{\sp}^4}+\frac{2p_0p_3^2}{P_{\sp}^6}\)\bigg],\\
\sumintf_{\{p_0\}} ~\frac{bp_3}{P_{\sp}^2}&=&-c_1 \sumintf_{\{p_0\}} \bigg[  \frac{p_3^2}{P_\sp^4}c_2+\frac{p_0p_3}{P_\sp^4}c_3-\bigg( \frac{p_3}{P_\sp^4}+\frac{2p_3^3}{P_\sp^6}\bigg)c_4 -\frac{4p_0p_3^2}{P_\sp^6} c_5\bigg],\\
\sumintf_{\{p_0\}} \frac{a^2}{P_{\sp}^2}&=&c_1^2 \sumintf_{\{p_0\}} \bigg[\frac{p_0^2}{P_\sp^6}c_2^2+\frac{p_3^2}{P_\sp^6}c_3^2+\frac{p_0^2p_3^2}{P_\sp^{10}}c_4^2+\bigg(\frac{1}{P_\sp^6}+\frac{4p_3^2}{P_\sp^8}+\frac{4p_3^4}{P_\sp^{10}} \bigg)c_5^2+\frac{2p_0p_3}{P_\sp^6}c_2 c_3\nn\\
&+&\frac{2p_0^2p_3}{P_\sp^8}c_2 c_4+\bigg( \frac{2p_0}{P_\sp^6}+\frac{4p_0p_3^2}{P_\sp^8}\bigg)c_2 c_5+\frac{2p_0p_3^2}{P_\sp^8} c_3 c_4+\bigg(\frac{2p_3}{P_\sp^6}+\frac{4p_3^3}{P_\sp^8} \bigg) c_3 c_5\nn\\
&+&\bigg(\frac{2p_0 p_3}{P_\sp^8}+\frac{4p_0p_3^3}{P_\sp^{10}} \bigg) c_4 c_5 \bigg],\\
\sumintf_{\{p_0\}} \frac{b^2}{P_{\sp}^2}&=&c_1^2\sumintf_{\{p_0\}}\bigg[\frac{p_3^2}{P_\sp^6}c_2^2+\frac{p_0^2}{P_\sp^6}c_3^2+\bigg( \frac{1}{P_\sp^6}+\frac{4p_3^2}{P_\sp^8}+\frac{4p_3^4}{P_\sp^{10}}\bigg)c_4^2+\frac{16p_0^2p_3^2}{P_\sp^{10}}c_5^2+\frac{2p_0p_3}{P_\sp^6}c_2 c_3\nn\\
&-&\bigg(\frac{2p_3}{P_\sp^6}+\frac{4p_3^3}{P_\sp^8}\bigg)c_2 c_4-\frac{8p_0p_3^2}{P_\sp^8}c_2 c_5-\bigg(\frac{2p_0}{P_\sp^6}+\frac{4p_0 p_3^2}{P_\sp^8} \bigg) c_3 c_4-\frac{8p_0^2p_3}{P_\sp^8} c_3 c_5\nn\\
&+&\bigg(\frac{8p_0 p_3}{P_\sp^8}+\frac{16p_0p_3^3}{P_\sp^{10}} \bigg) c_4 c_5\bigg],\\
\sumintf_{\{p_0\}} ~\frac{a^2p_3^2}{P_{\sp}^4}&=&c_1^2 \sumintf_{\{p_0\}} \bigg[\frac{p_0^2p_3^2}{P_\sp^8}c_2^2+\frac{p_3^4}{P_\sp^8}c_3^2+\frac{p_0^2p_3^4}{P_\sp^{12}}c_4^2+\bigg(\frac{p_3^2}{P_\sp^8}+\frac{4p_3^4}{P_\sp^{10}}+\frac{4p_3^6}{P_\sp^{12}} \bigg)c_5^2+\frac{2p_0p_3^3}{P_\sp^8}c_2 c_3\nn\\
&+&\frac{2p_0^2p_3^3}{P_\sp^{10}}c_2 c_4+\bigg( \frac{2p_0p_3^2}{P_\sp^8}+\frac{4p_0p_3^4}{P_\sp^{10}}\bigg)c_2 c_5+\frac{2p_0p_3^4}{P_\sp^{10}} c_3 c_4+\bigg(\frac{2p_3^3}{P_\sp^8}+\frac{4p_3^5}{P_\sp^{10}} \bigg) c_3 c_5\nn\\
&+&\bigg(\frac{2p_0 p_3^3}{P_\sp^{10}}+\frac{4p_0p_3^5}{P_\sp^{12}} \bigg) c_4 c_5 \bigg],\\
\sumintf_{\{p_0\}} ~\frac{b^2p_3^2}{P_{\sp}^4}&=&c_1^2\sumintf_{\{p_0\}}\bigg[\frac{p_3^4}{P_\sp^8}c_2^2+\frac{p_0^2p_3^2}{P_\sp^8}c_3^2+\bigg( \frac{p_3^2}{P_\sp^8}+\frac{4p_3^4}{P_\sp^{10}}+\frac{4p_3^6}{P_\sp^{12}}\bigg)c_4^2+\frac{16p_0^2p_3^4}{P_\sp^{12}}c_5^2+\frac{2p_0p_3^3}{P_\sp^8}c_2 c_3\nn\\
&-&\bigg(\frac{2p_3^3}{P_\sp^8}+\frac{4p_3^5}{P_\sp^{10}}\bigg)c_2 c_4-\frac{8p_0p_3^4}{P_\sp^{10}}c_2 c_5-\bigg(\frac{2p_0p_3^2}{P_\sp^8}+\frac{4p_0 p_3^4}{P_\sp^{10}} \bigg) c_3 c_4-\frac{8p_0^2p_3^3}{P_\sp^{10}} c_3 c_5\nn\\
&+&\bigg(\frac{8p_0 p_3^3}{P_\sp^{10}}+\frac{16p_0p_3^5}{P_\sp^{12}} \bigg) c_4 c_5\bigg],\\
\sumintf_{\{p_0\}} ~\frac{abp_0p_3}{P_{\sp}^4}&=&c_1^2\sumintf_{\{p_0\}}\bigg[-\frac{p_0^2p_3^2}{P_\sp^8}c_2^2-\frac{p_0^2p_3^2}{P_\sp^8}c_3^2+\bigg( \frac{p_0^2p_3^2}{P_\sp^{10}}+\frac{2p_0^2p_3^4}{P_\sp^{12}}\bigg)c_4^2+\bigg(\frac{4p_0^2p_3^2}{P_\sp^{10}}+\frac{8p_0^2p_3^4}{P_\sp^{12}} \bigg)c_5^2\nn\\
&-&\bigg( \frac{p_0^3 p_3}{P_\sp^8}+\frac{p_0 p_3^3}{P_\sp^8}\bigg)c_2c_3+\bigg(\frac{p_0^2p_3}{P_\sp^8}+\frac{p_0^2 p_3^3}{P_\sp^{10}} \bigg)c_2c_4+\bigg(\frac{4p_0^3p_3^2}{P_\sp^{10}}-\frac{p_0p_3^2}{P_\sp^8}-\frac{2p_0p_3^4}{P_\sp^{10}} \bigg) c_2 c_5\nn\\
&+&\bigg(\frac{p_0p_3^2}{P_\sp^8}-\frac{p_0^3p_3^2}{P_\sp^{10}}+\frac{2p_0p_3^4}{P_\sp^{10}} \bigg)c_3c_4+\bigg(\frac{2p_0^2 p_3^3}{P_\sp^{10}}-\frac{p_0^2p_3}{P_\sp^8} \bigg) c_3c_5\nn\\
&+&\bigg( \frac{p_0p_3}{P_\sp^8}+\frac{4p_0p_3^3}{P_\sp^{10}}+\frac{p_0^3p_3^3}{P_\sp^{12}}+\frac{4p_0p_3^5}{P_\sp^{12}}\bigg)c_4c_5\bigg],
\eea
which leads to
\bea
F'_q&=& -4 d_F \sum_f \frac{q_fB}{(2\pi)^2}\sumintf_{\{p_0\}}\Bigg[\frac{c_1 c_2}{P_\sp^2}+\bigg(\frac{2p_3}{P_\sp^4}+\frac{3p_3^3}{P_\sp^6}\bigg)c_1 c_4 +\bigg( \frac{p_0}{P_\sp^4}+\frac{6p_0p_3^2}{P_\sp^6}\bigg) c_1 c_5 -\frac{c_1^2c_2^2}{P_\sp^4}  -\frac{c_1^2c_3^2}{P_\sp^4}\nn\\
&-&\bigg(\frac{1}{P_\sp^6}+\frac{11p_3^2}{P_\sp^8}+\frac{27p_3^4}{P_\sp^{10}}+\frac{18p_3^6}{P_\sp^{12}} \bigg)c_1^2 c_4^2 -\bigg(\frac{1}{P_\sp^6}+\frac{38p_3^2}{P_\sp^8}+\frac{108p_3^4}{P_\sp^{10}}+\frac{72p_3^6}{P_\sp^{12}} \bigg)c_1^2c_5^2\nn\\
&-&\bigg( \frac{4p_3}{P_\sp^6}+\frac{6p_3^3}{P_\sp^8}\bigg)c_1^2 c_2 c_4-\bigg(\frac{2p_0}{P_\sp^6}+\frac{12p_0p_3^2}{P_\sp^8} \bigg)c_1^2 c_2 c_5+\bigg( \frac{2p_0}{P_\sp^6}+\frac{6p_0p_3^2}{P_\sp^8}\bigg)c_1^2c_3c_4\nn\\
&+&\bigg( \frac{10p_3}{P_\sp^6}+\frac{12p_3^3}{P_\sp^8}\bigg)c_1^2c_3c_5-\bigg( \frac{14p_0p_3}{P_\sp^8}+\frac{72p_0p_3^3}{P_\sp^{10}}+\frac{72p_0p_3^5}{P_\sp^{12}}\bigg)c_1^2c_4c_5\Bigg]\nn\\
&=&-4 d_F \sum_f \frac{q_fB}{(2\pi)^2}\sumintf_{\{p_0\}}\Bigg[\frac{c_1 c_2}{P_\sp^2}-\frac{c_1^2c_2^2}{P_\sp^4}  -\frac{c_1^2c_3^2}{P_\sp^4}-\bigg(\frac{1}{P_\sp^6}+\frac{11p_3^2}{P_\sp^8}+\frac{27p_3^4}{P_\sp^{10}}+\frac{18p_3^6}{P_\sp^{12}} \bigg)c_1^2 c_4^2 \nn\\
&-&\bigg(\frac{1}{P_\sp^6}+\frac{38p_3^2}{P_\sp^8}+\frac{108p_3^4}{P_\sp^{10}}+\frac{72p_3^6}{P_\sp^{12}} \bigg)c_1^2c_5^2\Bigg].\label{quark_fe}
\eea

One can calculate the Matsubara frequency sum as~\cite{Bellac:2011kqa}
\bea
 \sum_{\{p_0\}} \frac{1}{P_{\sp}^2}&=& -\frac{1}{2 p_3}\bigg(1-  n_F(p_3+\mu)-n_F(p_3-\mu)\bigg)\label{P2_sum_F},
\eea
The first term in Eq.~\eqref{P2_sum_F} is temperature independent vacuum term and can be regularized using regular vacuum regularization. The regularized contribution will be temperature independent and  will not contribute anything to the thermodynamics of the system. So, without vacuum term, Eq.~\eqref{P2_sum_F} becomes
\bea
\sum_{\{p_0\}} \frac{1}{P_{\sp}^2}&=& \frac{1}{2p_3}\bigg(n_F(p_3+\mu)+n_F(p_3-\mu) \bigg).
\label{P2_sum_F2}
\eea

Now one can write the sum-integral as
\bea
\sum\!\!\!\!\!\!\!\!\int \frac{1}{P_{\sp}^2}&=&\(\frac{e^{\gamma_E}\Lambda^2}{4\pi }\)^{\epsilon}\int_{-\infty}^{\infty}\frac{d^{1-2\eps}p_3}{2p_3} \Big(n_F(p_3+\mu)+n_F(p_3-\mu)\Big).
\eea

We perform the sum-integrals~\cite{Karmakar:2019tdp} in Eq.~\eqref{quark_fe} as
\bea
\sumintf_{\{p_0\}} \, \frac{1}{P_{\sp}^{2}}&=&\(\frac{e^{\gamma_E}\Lambda^2}{4\pi }\)^{\epsilon}\int_{-\infty}^{\infty} d^{1-2\eps}p_3\, \frac{n_F(p_3)}{p_3}\nn\\
&=&\(\frac{\Lambda}{4\pi T}\)^{2\eps}\Bigg[-\frac{1}{2\eps}-\frac{1}{2}\(3\gamma_E+4\ln{2}-\ln{\pi}\)+\frac{7\mu^2\zeta(3)}{4\pi^2T^2} -\frac{31\mu^4\zeta(5)}{16\pi^4T^4}+\mathcal O(\eps)\Bigg],\\
\sumintf_{\{p_0\}} \, \frac{1}{P_{\sp}^4}
&=&\(\frac{\Lambda}{4\pi T}\)^{2\eps}\Bigg[\frac{7}{8\pi^2} \frac{\zeta (3)}{T^2}-\frac{93\mu^2\zeta(5)}{16\pi^4T^4}31+\frac{1905\mu^4\zeta(7)}{128\pi^6T^6}+\mathcal O(\epsilon)\Bigg],\\
  \sumintf_{\{p_0\}} \, \frac{1}{P_{\sp}^{6}}
  &=&\(\frac{\Lambda}{4\pi T}\)^{2\eps}\Bigg[-\frac{93\,\zeta(5)}{128\pi^4 T^4}+\frac{5715\mu^2\zeta(7)}{512\pi^6T^6}-\frac{53655\mu^4\zeta(9)}{1024\pi^8 T^8}+\mathcal O(\eps)\Bigg],\\
   \sumintf_{\{p_0\}} \, \frac{p_3^2}{P_{\sp}^{8}}
   &=&\(\frac{\Lambda}{4\pi T}\)^{2\eps}\Bigg[\frac{31\,\zeta(5)}{256 \pi^4 T^4}-\frac{1905\mu^2\zeta(7)}{1024\pi^6T^6}+\frac{17885\mu^4\zeta(9)}{2048\pi^8T^8}+\mathcal O(\eps)\Bigg],\\
    \sumintf_{\{p_0\}} \, \frac{p_3^4}{P_{\sp}^{10}}
   &=& \(\frac{\Lambda}{4\pi T}\)^{2\eps}\Bigg[-\frac{93\zeta(5)}{2048\pi^4T^4}31+\frac{5715\mu^2\zeta(7)}{8192\pi^6T^6}-\frac{53655\mu^4\zeta(9)}{16384\pi^8T^8}+\mathcal O(\eps)\Bigg],\\
    \sumintf_{\{p_0\}} \, \frac{p_3^6}{P_{\sp}^{12}}
   &=&\(\frac{\Lambda}{4\pi T}\)^{2\eps}\Bigg[\frac{93\zeta(5)}{4096\pi^4T^4}-\frac{5715\mu^2\zeta(7)}{16384\pi^6T^6}+\frac{53655\mu^4\zeta(9)}{32768\pi^8T^8}+\mathcal O(\eps)\Bigg],
\eea
   where $\Lambda$ is the ${\overline {\mbox{MS}}}$ renormalization scale.

Using the above sum-integrals in Eq.~(\ref{quark_fe}) $F'_q$ up to $\mathcal{O}(g^4)$ becomes,

\bea
F'_q&=&-4 d_F \sum_f \frac{q_fB}{(2\pi)^2}\sumintf_{\{p_0\}}\Bigg[\frac{c_1 c_2}{P_\sp^2}-\frac{c_1^2c_2^2}{P_\sp^4}  -\frac{c_1^2c_3^2}{P_\sp^4}-\bigg(\frac{1}{P_\sp^6}+\frac{11p_3^2}{P_\sp^8}+\frac{27p_3^4}{P_\sp^{10}}+\frac{18p_3^6}{P_\sp^{12}} \bigg)c_1^2 c_4^2 \nn\\
&-&\bigg(\frac{1}{P_\sp^6}+\frac{38p_3^2}{P_\sp^8}+\frac{108p_3^4}{P_\sp^{10}}+\frac{72p_3^6}{P_\sp^{12}} \bigg)c_1^2c_5^2\Bigg]\nn\\
&=&4 d_F \sum_f \frac{q_fB}{(2\pi)^2}\frac{g^2 C_F (q_f B)}{4 \pi^2}\(\frac{\Lambda}{4\pi T}\)^{2\eps}\Bigg[\frac{1}{\eps}\bigg(-\frac{1}{2}\ln 2+\frac{7\mu^2 \zeta(3)}{16\pi^2 T^2}-\frac{31\mu^4 \zeta(5)}{64\pi^4T^4} \bigg)-\frac{3\gamma_E \ln 2}{2}\nn\\
&+&\ln 2 \ln \pi-\frac{1}{2}\ln 2 \ln 16\pi +\frac{g^2 C_F (q_f B)}{4 \pi^2}\frac{63\ln2^2 \zeta(3)}{72\pi^2T^2}-\frac{g^2 C_F (q_f B)}{4 \pi^2}\frac{217(q_fB)^2\zeta(5)}{36864\pi^4T^6}\nn\\
&\times&\Big(\gamma_E+2\ln2-12\ln G \Big)^2+\frac{\mu^2}{1152\pi^2T^4}\bigg\{ \frac{7\zeta(3) g^2 C_F (q_f B)}{4 \pi^2}\Big( 3+3\gamma_E+4\ln 2 -36\ln G\Big)^2\nn\\
&+&504T^2\zeta(3)\Big(3\gamma_E+8\ln 2-\ln \pi \Big)-\frac{36 \ln 2}{\pi^2}\frac{g^2 C_F (q_f B)}{4 \pi^2}\Big( 49\zeta(3)^2+186\ln 2 \zeta(5)\Big)\bigg\}\nn\\
&-&\frac{7 (q_fB)^2\mu^2}{737280\pi^4T^8}\frac{g^2 C_F (q_f B)}{4 \pi^2}\bigg\{-31\zeta(5)\Big( -15+15\gamma_E+16\ln 2\Big)\Big( \gamma_E+2\ln 2 -12\ln G\Big)\nn\\
&-&\frac{48825}{\pi^4}\zeta(3)^2\zeta(5)-\frac{9525\zeta(7)}{\pi^2}\Big(\gamma_E+2\ln 2-12\ln G \Big)^2+55800\zeta(5)\zeta'(-3)\nn\\
&\times&\Big(\gamma_E+2\ln2-12\ln G \Big) \bigg\}+\frac{\mu^4}{69120\pi^6 T^6}\bigg\{-1080\pi^2T^2\bigg(98\zeta(3)^2+31\zeta(5)\Big( 3\gamma_E+8\ln 2 -\ln \pi\Big) \bigg)\nn\\
&-&\frac{g^2 C_F (q_f B)}{4 \pi^2}\bigg(14\pi^4 \zeta(3)\Big( 15\gamma_E+16\ln 2\Big)\Big( 3+3\gamma_E+4\ln 2-36\ln G\Big)-46305\zeta(3)^3\nn\\
&+&2790\pi^2 \zeta(5)\Big( 3+3\gamma_E-4\ln 2-36\ln G\Big)^2-820260 \ln 2 \zeta(3) \zeta(5)-1028700\ln 2^2 \zeta(7)\nn\\
&-&25200\pi^4 \zeta(3) \zeta'(-3)\Big(3+3\gamma_E+4\ln 2-36\ln G \Big) \bigg) \bigg\}-\frac{ \mu^4 (q_fB)^2}{5308416\pi^4 T^{10}}\frac{g^2 C_F (q_f B)}{4 \pi^2}\nn\\
&\times&\bigg\{ \frac{10897740}{\pi^6}\zeta(3)\zeta(5)^2+\frac{37804725}{\pi^6}\zeta(3)^2 \zeta(7)+\frac{2253510\zeta(9)}{\pi^4}\Big( \gamma_E+2\ln 2-12\ln G\Big)^2\nn\\
&+&\frac{24003}{\pi^2}\zeta(7)\Big( \gamma_E+2\ln 2-12\ln G\Big)\Big(-15+15\gamma_E+16\ln 2-1800\zeta'(-3) \Big)\nn\\
&+&\frac{31\zeta(5)}{100}\bigg(14175-40950\gamma_E+26775\gamma_E^2+68240\gamma_E \ln 2+41728\ln 2^2+151200\ln G\nn\\
&-&151200\gamma_E \ln G-240\ln 2 \Big( 231+640\ln G\Big)+3175200\zeta'(-5)\Big( \gamma_E+2\ln 2-12\ln G\Big)\nn\\
&-&226800\zeta'(-3)\Big( -15+15\gamma_E+16\ln 2\Big)+204120000\zeta'(-3)^2\bigg)\bigg\}\Bigg]
\eea


\end{document}